%% file: main.tex
\documentclass[journal = jacsat, manuscript = article]{achemso}
\input{definitions.tex}
\input{head.tex}
\usepackage{hyperref}
\SectionNumbersOn
\begin{document}
\maketitle

\begin{abstract}
We present the explicit bonding Reaction ensemble Monte Carlo (eb-RxMC) method, designed to sample reversible bonding reactions in macromolecular systems in thermodynamic equilibrium.
Our eb-RxMC method is based on the Reaction ensemble method, however, its implementation differs from the latter by the representation of the reaction.
In the eb-RxMC implementation, we are adding or deleting bonds between existing particles, instead of inserting or deleting particles with different chemical identities.
This new implementation makes the eb-RxMC method suitable for simulating the formation of reversible linkages between macromolecules, which would not be feasible with the original implementation.
To enable coupling of our eb-RxMC algorithm with Molecular Dynamics algorithm for sampling of the configuration space, we biased the sampling of reactions only within a certain inclusion radius.
We validated our algorithm using a set of ideally behaving systems undergoing dimerization and polycondensation reactions, for which analytical results are available.
For dimerization reactions with various equilibrium constants and initial compositions, the degree of conversion measured in our simulations perfectly matched the reference values given by the analytical equations.
We also showed that this agreement is not affected by the arbitrary choice of the inclusion radius or the stiffness of the harmonic bond potential.
Next, we showed that our simulations can correctly match the analytical results for the distribution of the degree of polymerization and end-to-end distance of ideal chains in polycondensation reactions.
Altogether, we demonstrated that our eb-RxMC simulations correctly sample both reaction and configuration space of these reference systems, opening the door to future simulations of more complex interacting macromolecular systems.
\end{abstract}

\section{Introduction}

Reversible chemical reactions are often present in macromolecular systems.
Metathesis and polycondensation are typical examples of reversible reactions reaching equilibrium during the polymer synthesis.\cite{Flory1969,Flory1936,Munk1989,Hocker1991}
These polymerization reactions typically lead to a broad distribution of degrees of polymerization\cite{Keki2013,Gavrilov2017,lang2021a,lang2021b} sometimes also leading to crosslinking and gelation.\cite{Rudyak2019,Gavrilov2020,Nian2023}
The self-healing properties of polymer materials often rely on reversible formation of crosslinks, which can be easily re-formed after they have been broken due to an external stimulus.\cite{stukalin13a,Parada2018,Othman2022}
The association of boronic acids is an interesting example of such a reaction which competes with the acid-base equilibrium reactions on the same functional groups.
It has been hypothesized that crosslinking between polymers functionalized with boronic acids is responsible for peculiar properties of these polymers but the actual mechanism of this crosslinking remains unclear.\cite{Vdorvdovivc2019,Markova2021}
Chelation of metals to polyelectrolyte chains is known to collapse polymer chains\cite{Koper2001,Nap2018}, sometimes also leading to crosslinking and gelation.\cite{Ermoshkin2003}
Last but not least, the reversible formation of hydrogen bonds plays a crucial role in many biochemical and biological systems, as well as in synthetic polymers.\cite{Khutoryanskiy2009}

The common feature of the reversible bonds described  above is that they connect two different macromolecules, or two parts of the same macromolecule, which are close in space but can be far from each other when measured along the contour of the macromolecule.
On the one hand, the probability of forming these bonds is determined by the chemical nature of the reactants, which can be phenomenologically described by the equilibrium constant $K$ of the given chemical reaction.
On the other hand, the spatial distribution of the reacting groups is determined not only by their overall concentration but also by the conformational properties of the macromolecule.
This coupling between the conformation and reversible bond formation poses a fundamental challenge for modelling of such systems.
Consequently, theoretical description of such reversible reactions are typically limited to analytical theories which assume homogeneous distribution of the reactive groups.
Molecular simulation approaches fully accounting for the coupling between the reversible bond formation (reaction space) and conformation of the macromolecules (configuration space) are very scarce.

Analytical theories describing reaction equilibria in macromolecular systems usually rely on strong approximations, which limit their applicability, while simultaneously writing the results as a closed-form analytical expression.
In the ideal gas limit, the equilibrium composition of a reacting mixture can be analytically determined from the equilibrium constant $K$ of the reactions and the initial composition of the system.
In the case of linear polycondensation reactions, the classical theory from Flory\cite{Flory1936} permits calculating the molecular weight distribution of the polymer chains at a given degree of conversion of the reaction $p$.
More modern theories include cyclization of the polymer chains, allowing to calculate the fraction of linear chains and the fraction of rings.\cite{lang2021a,lang2021b}
In general, these theories can be augmented beyond the ideal gas approximation by accounting for the excess chemical potential or, equivalently, the activity of the reactants.
Such an approach has been used for example in the analytical theories describing acid-base equilibria in weak polyelectrolytes, accounting in an approximate manner for coupling between the reaction and configuration space of the macromolecule.\cite{Garces2014,Garces2017,Blanco2018}
However, accounting for non-ideal behaviour due to interactions in reactive macromolecular systems is beyond the capabilities of many analytical theories.

As an alternative to analytical theories, computer simulations can numerically  sample the configuration space and simultaneously the reaction space.
The full description of chemical reactions requires the use of quantum mechanics to determine the electronic structure, which is not feasible for macromolecular systems due to high computational cost.
The use of hybrid Quantum Mechanics / all-atom Molecular Dynamics slightly alleviates the problem but remains restricted to a small number of reactive chemical species.\cite{Mieres2020}
The computational cost can be further mitigated, by replacing the Quantum description of the chemical reactions by effective reactive potentials.\cite{paciolla2022,carvalho2022}
However, the above methods are  computationally too demanding for macromolecular systems which include long chains and hundreds of reactive groups simultaneously participating in the equilibrium reaction.
At this level, a further reduction of the resolution of the model is required, representing whole functional groups or monomeric units  as single particles, referred to as coarse-grained  models.
At this level, chemical reactions can be treated as stochastic events and simulated using Monte Carlo (MC) methods.
The reaction is then represented by a change of chemical identities of the coarse-grained particles, or by the formation or disappearance of chemical bonds between existing particles.
In such phenomenological description, the thermodynamic equilibrium constant and the reaction stoichiometry are the only input parameters, whereas the detailed reaction mechanism becomes irrelevant.
Indeed, MC simulations in various statistical ensembles are now commonly used to study pH-sensitive polymer systems.\cite{landsgesell19a,blanco2023} 
These simulations are restricted to reactions which involve a change of chemical identities, while simultaneously releasing a small particle from a macromolecule or attaching this particle in the reverse step of the reaction.
However, such a representation cannot be smoothly extended to represent reversible bond formation between two macromolecules or even oligomers.

The Reaction ensemble Monte Carlo (RxMC) \cite{johnson1994,smith1994} method offers a general framework for sampling chemical equilibria.
In the original implementation of the RxMC, the chemical reaction is represented by an exchange of particles with a virtual reservoir.
Smith and Triska\cite{smith1994} validated the RxMC method using reversible dimerization reactions in which two reactant molecules associate to create a product molecule. 
In their implementation, a forward direction of the dimerization reaction entails deleting two reactant particles from the system and inserting a new product particle into the system.
The reverse direction entails deleting the product particle and inserting two reactant particles.
In this description, the formation and breaking of chemical bonds is represented only \emph{implicitly} through the equilibrium constant $K = \exp( -\Delta_{\mathrm{r}}G / RT)$, where $\Delta_{\mathrm{r}}G$ is the standard Gibbs free energy of reaction. 
When applied to macromolecules, such implicit description of the bonding is sufficient only if it represents binding of small reactants to big molecules, for example when studying the protonation / deprotonation of weakly acidic or basic groups. \cite{uhlik14a,nova17a,landsgesell19a,blanco2023}
However, in other cases, it is desirable to implement the chemical reaction as the creation or deletion of bonds between particles which already exist in the simulated system.
Examples of such cases include cross-linking reactions in hydrogels,\cite{stukalin13a,Parada2018,Othman2022} chelation of metals in polyelectrolytes\cite{Koper2001,Ermoshkin2003,Nap2018}  or various polymerization reactions.\cite{Keki2013,Gavrilov2017,lang2021a}. 

Examples of simulations of reactive macromolecular systems with explicit reversible bonds are relatively scarce in the literature.
Radical polymerization reactions have been studied with reactive MC schemes in which permanent bonds are created between nearby pairs of reactive particles. \cite{Gavrilov2017,Gavrilov2020}
Polycondensation reactions with cyclization have been simulated using MC schemes with probability criteria based on kinetic arguments.\cite{lang2021a,lang2021b}
A modified RxMC scheme, including explicit description of the reversible bonds, has been proposed to investigate polymerization reactions in silica gels. \cite{malani2010,malani2011,chien2015} 
The challenge in these simulations is to simultaneously sample the reaction and configuration spaces of the macromolecules.
In a pure MC approach, the reactive MC moves can be combined with configuration moves.\cite{malani2010,malani2011,chien2015} 
However, these configuration moves need to be carefully chosen and fine-tuned for each specific case to enable efficient sampling of the system.\cite{Oyarzun2018a,Oyarzun2018b}
Alternatively, one can combine the reactive MC moves with Molecular Dynamics (MD) to explore the configuration space. 
The advantage of such a combination is that the MD part uses the collective dynamics of the whole system to sample the configuration space, which is both universal and reasonably efficient also for dense systems.
Simultaneously, the combination of MC with MD imposes additional restrictions.
In such a combined approach, one typically needs to restrict the reaction MC moves to only select pairs of neighboring particles.
Otherwise, occasional acceptance of energetically less favourable configurations would result in big forces due to the newly created bonds, which could break a subsequent MD integration scheme. \cite{Gavrilov2017,Gavrilov2020}
Therefore, combining the MC schemes for sampling the reactions with MD integration for sampling the system configurations remains a challenge.

\begin{figure}[H]
\centering
\includegraphics [ width =0.99\textwidth]{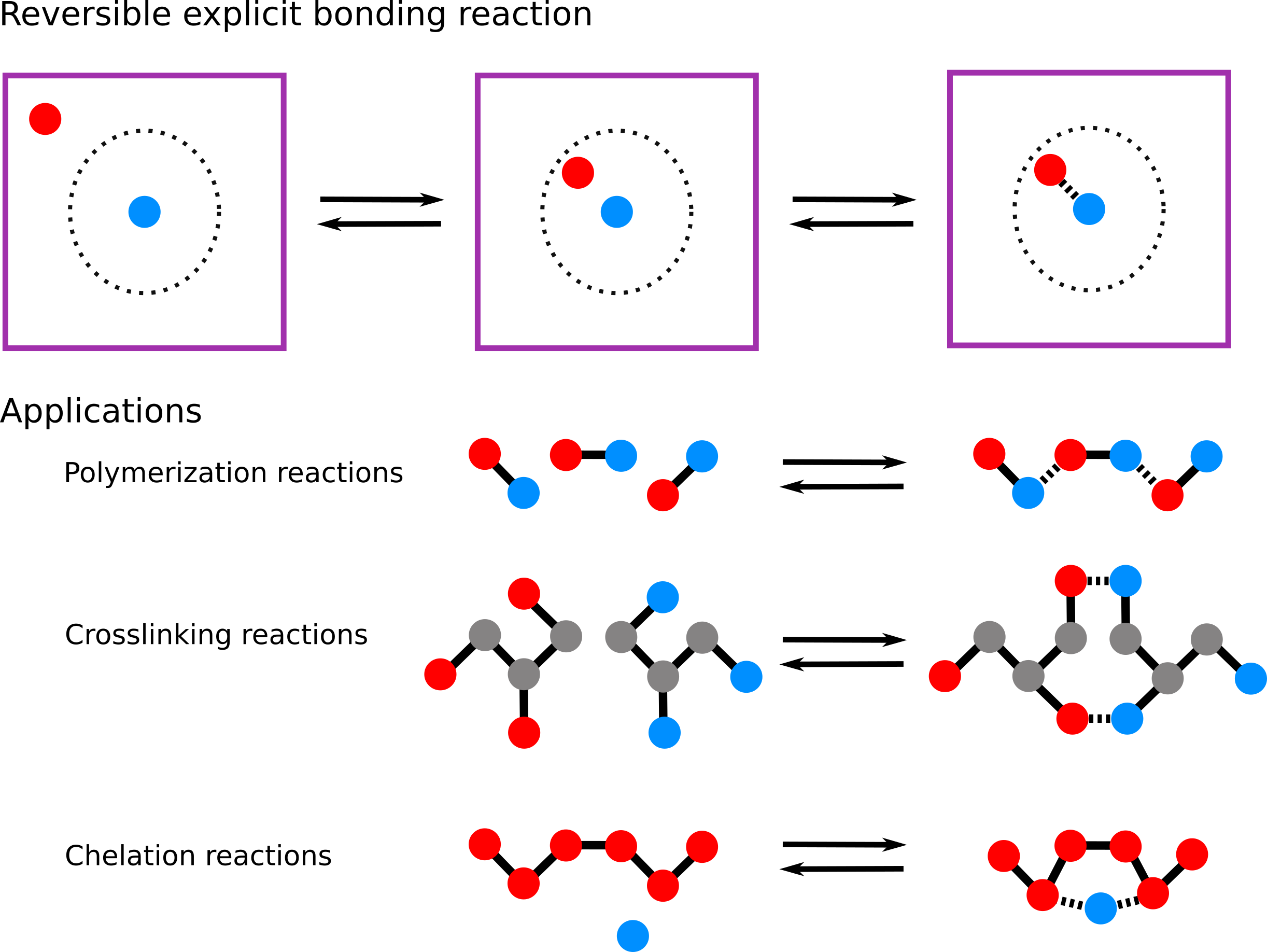}
\caption{Top panel: Scheme illustrating our proposed explicit bonding Reaction Ensemble Monte Carlo (eb-RxMC) algorithm that creates explicit reversible bonds between particles. These particles can only react when they are at a distance below an inclusion radius (dashed circle).
A suitably chosen inclusion radius permits coupling of the eb-RxMC algorithm with Molecular Dynamics. 
Bottom panel: schematic examples of reactions which can be simulated using our proposed algorithm. 
In the illustrations, permanent bonds are represented by solid lines whereas reversible bonds are represented by dashed lines.
In all panels blue and red particles represent reactive chemical groups that can react with each other while grey circles represent chemically inert groups.}
\label{fig:general_scheme}
\end{figure}

In this work, we present a biased MC algorithm to sample chemical equilibrium in the Reaction ensemble including \emph{explicit} formation of reversible bonds.
We refer to this algorithm as explicit bonding Reaction Ensemble Monte Carlo (eb-RxMC) to
distinguish it from the original implementation, in which the reaction is modeled by inserting and deleting particles from the simulation box.\cite{johnson1994,smith1994}
Our eb-RxMC can be constrained to only sample the reaction within an inclusion radius $\rin$ which permits combining it with MD schemes, as illustrated in Fig. \ref{fig:general_scheme} (upper panel).
The intended use case of the eb-RxMC method are simulations of reversible bonding reactions in macromolecular systems in thermodynamic equilibrium, schematically shown in Fig. \ref{fig:general_scheme} (lower panel). 
They include, but are not limited to, the following types of reactions:
polymerization reactions reaching equilibrium (\eg polycondensation reactions),\cite{Keki2013,lang2021a,lang2021b}
reversible cross-linking reactions,\cite{stukalin13a,Parada2018,Othman2022}
reversible non-covalent bonds (\eg hydrogen bonding),\cite{Vdorvdovivc2019,Markova2021}
chelation of metal ions in macromolecular systems\cite{Koper2001,Ermoshkin2003,Nap2018} and polyassociation reactions in dynamers.\cite{Lehn2005}
Our eb-RxMC is designed to sample thermodynamic equilibrium properties of reversible bonding reactions, avoiding the need to describe the actual reaction mechanism or to do any specific assumption on the kinetics of the reaction.
Therefore, sampling of the chemical reaction proceeds at a user-defined rate, allowing the user to control the effective time scale on which the reaction equilibrium is established.
As a downside, this means that our eb-RxMC is not designed to provide any information about the actual dynamics of the system.
In return, our eb-RxMC is able to efficiently sample the equilibrium properties of macromolecular systems with several hundreds of reactive groups. 

In the following sections, we first provide an overview of some known analytical results for dimerization and polycondensation reactions which we use to establish the nomenclature and as benchmarks to validate our eb-RxMC method.
Next, we describe our implementation of the eb-RxMC method and how we bias it to enable its coupling with Langevin Dynamics.
Then, we validate our eb-RxMC algorithm for a set of ideally behaving systems, using the analytical results presented in the previous sections.
We conclude by outlining several interesting open problems that could be addressed by our eb-RxMC method in the near future, indicating probable further development of the method.

\section{Theoretical background}

In this section, we recapitulate some known analytical results for the equilibrium degree of conversion of dimerization and polycondensation reactions.
This recapitulation is provided in order to introduce a unified notation for both types of reactions.
These analytical results will be used later as a benchmark for testing the numerical results of our eb-RxMC simulations.

\subsection{Equilibrium in dimerization reactions}

The dimerization reaction can involve either association of two identical molecules (monomers) or two different ones, schematically represented by the reactions
\begin{subequations}
  \begin{equation}
    \label{eq:AAbonding}
      \ce{A + A <=> A-A}
  \end{equation}
  \begin{equation}
    \label{eq:ABbonding}
    \ce{A + B <=> A-B}
  \end{equation}
\end{subequations}
where A and B represent different functional groups which can react with each other by forming a reversible bond.

The chemical equilibrium of the reversible bonding reactions is defined by the equilibrium constant $K$, related to the Gibbs free energy of the reaction as
\begin{equation}
    \label{eq:DeltaG_K}
    \Delta_{\mathrm{r}} G = - RT \ln K
\end{equation}
where $T$ is the temperature and $R$ is the molar gas constant.
The equilibrium constant of Eq. \ref{eq:AAbonding} can be expressed in terms of activities of the reacted groups \ce{A-A} and free groups \ce{A} and analogically, equilibrium constant of Eq. \ref{eq:ABbonding} can be expressed in terms of activities of the reacted groups \ce{A-B} and free groups \ce{A} and \ce{B}:
\begin{subequations}
\begin{equation}
K \equiv  \frac{ a_{\ce{A-A}} }{a_{\ce{A}}^2}
\overset{\mathrm{ideal}}{=} \frac{c_{\ce{A-A}} \, \cref}{c_{\ce{A}}^2}
\,,
\label{eq:K_AA_id}
\end{equation}
\begin{equation}
K \equiv  \frac{ a_{\ce{A-B}} }{a_{\ce{A}} a_{\ce{B}}}
\overset{\mathrm{ideal}}{=} \frac{c_{\ce{A-B}} \, \cref}{c_{\ce{A}}c_{\ce{B}}}
\,,
\label{eq:K_AB_id}
\end{equation}
\end{subequations}
where $c^\ominus$ is the reference concentration, conventionally chosen as $c^{\ominus} = \qty{1}{mol/kg}$.\cite{mcnaught1997}
In the ideal gas limit, {\ie} in absence of interactions, the activities of various functional groups can be replaced with their concentrations, as indicated by the second equality in  Eqs. \ref{eq:K_AA_id} and  \ref{eq:K_AB_id}.
For reaction Eq. \ref{eq:AAbonding}, we define the degree of conversion as
\begin{equation}
p \equiv \frac{c_{\ce{A-A}}}{\cmax_{\ce{A-A}}} = \frac{ \ctot -c_{\ce{A}}}{\ctot},
\label{eq:p_AA}
\end{equation}
where $\cmax_{\ce{A-A}}$ is the maximum attainable concentration of \ce{A-A} bonds and $\ctot$ is the total concentration of reactive groups, i.e. $\ctot=c_{\ce{A}} + 2c_{\ce{A-A}}$.
For reaction Eq. \ref{eq:ABbonding} we define the degree of conversion in an analogous manner
\begin{equation}
p \equiv \frac{c_{\ce{A-B}}}{\cmax_{\ce{A-B}}} = \frac{\ctot -c_{\ce{A}} -c_{\ce{B}}}{\ctot (1-\rex)},
\label{eq:p_AB}
\end{equation}
where $\cmax_{\ce{A-B}}$ is the maximum attainable concentration of \ce{A-B} bonds, $\ctot$ is the total concentration of reactive groups, i.e. $\ctot=c_{\ce{A}} + c_{\ce{B}} + 2c_{\ce{A-B}}$ and $\rex$ is the excess ratio, defined as
\begin{equation}
\rex = | 2f_{\ce{A}} -1 |
\label{eq:rex}
\end{equation}
where $f_{\ce{A}}$ is the mole fraction of A groups in the reactive mixture, including both bonded and free A groups.
The excess ratio characterizes deviations of the composition of the reactive mixture from the ideal equimolar ratio of A and B.

In the ideal limit, the degree of conversion at equilibrium can be expressed by the following equations:
\begin{subequations}
  \begin{equation}
    p = 1 + \frac{1-\sqrt{1+8 \lambda}}{4 \lambda}
    \qquad\text{(Eq. \ref{eq:AAbonding}, ideal gas),}
    \label{eq:p_AA_id}
  \end{equation}
  \begin{equation}
    p =  (1-r_{\ce{ex}})^{-1} \left(1+ \frac{1-\sqrt{1 + 2\lambda+\lambda^2r_{\ce{ex}}^2}}{\lambda}\right) \qquad\text{(Eq. \ref{eq:ABbonding}, ideal gas),}
    \label{eq:p_AB_id}
  \end{equation}
\end{subequations}
where $\lambda = K \ctot / \cref$. 
A detailed derivation of Eqs. \ref{eq:p_AA_id} and \ref{eq:p_AB_id} is provided in the Supporting Information (Section S1).
Typically, one needs values of $K$ or $\ctot$ spanning over various orders of magnitude to cover the whole range of possible $p$ values.
Therefore, it is convenient to introduce $\plambda$, defined as
\begin{equation}
\plambda = -\log_{10}(\lambda) = \pK + \pC
\label{eq:plambda}
\end{equation}
where $\pK=-\log_{10}(K)$ and $\pC = -\log_{10}(\ctot/\cref)$. This logarithmic scale is in analogy with the acidity constant $\pK_\mathrm{A}=-\log_{10}(K_\mathrm{A})$, commonly used to describe the acid-base equilibria.\cite{landsgesell19a,blanco2023}

\subsection{Equilibrium in polycondensation reactions}
If the molecule (monomer) contains two reactive groups, then it can form polymer chains in a polycondensation reaction
\begin{subequations}
\begin{equation}
\ce{A-A + A-A} \ce{<=> (A-A)_2}
\label{eq:l_dimerization}
\end{equation}
\begin{equation}
\ce{(A-A)_n + (A-A)_m} \ce{<=> (A-A)_{n+m} }
\label{eq:l_polymerization}
\end{equation}
\end{subequations}
where $n$ and $m$ are integers, referring to the number of \ce{A-A} monomers comprising the chain molecule, i.e. the degree of polymerization of the chain.
The degree of conversion for Eq. \ref{eq:l_polymerization} can be defined in the same way as Eq. \ref{eq:p_AA} and its value at equilibrium is given by Eq. \ref{eq:p_AA_id}.
Flory\cite{flory36a} assumed that $\Delta_{\mathrm{r}} G$ of reactions Eqs. \ref{eq:l_dimerization} and \ref{eq:l_polymerization} does not depend on the values of $m$ and $n$.
Under this assumption, he obtained an analytical expression for the mass fraction of chains composed of $M$ monomers
\begin{equation}
    w(M) = Mp^{M-1}(1-p)^2.
    \label{eq:Flory}
\end{equation}
The mass fraction  is defined as
\begin{equation}
    w(M) = \frac{ M N_{\mathrm{M}}}{M_{\mathrm{t}}},
\end{equation}
were $N_{\mathrm{M}}$ is the number of chains consisting of $M$ monomers and $M_{\mathrm{t}} = \sum_{M} M N_M$ is the total number of monomers in solution.

\section{Method and model\label{sec:method}}

In our simulation model, we employ a simplified coarse-grained representation, where one particle in the simulation represents one functional group of a molecule, as defined in Eqs. \ref{eq:AAbonding}, \ref{eq:ABbonding}, \ref{eq:l_dimerization} and \ref{eq:l_polymerization}.
We simulate an ensemble of these molecules, evolving in the reaction space, \ie, undergoing the chemical reactions, and simultaneously evolving in the configuration space, \ie, undergoing conformational changes and diffusing in the simulation box.
The evolution in reaction space is simulated using the explicit bonding reaction ensemble Monte Carlo (eb-RxMC).
The evolution in configuration space is simulated using Langevin Dynamics (LD) in implicit solvent.
One simulation cycle thus consists of a set of eb-RxMC moves, followed by a set of integration steps in the LD.
The whole simulation then comprises many of such cycles.
Implementation of the eb-RxMC protocol which enables its coupling to LD comprises the key result of our work.

\subsection{The explicit bonding Reaction Monte Carlo algorithm \label{sec:RxMC}}
Our eb-RxMC implementation consists of the following steps, which are described in more detail below:
1. Select a forward or reverse direction of the reaction and select the reactive groups involved; 2. Compute the acceptance probability; 3. Change to a new state if the trial move is accepted, \ie  create or delete bonds between the reactive particles in the simulation box, mimicking Eq. \ref{eq:AAbonding} or Eq. \ref{eq:ABbonding}.
These trial moves are accepted with a probability given by the Reaction ensemble. \cite{johnson1994,smith1994} 
We have implemented the eb-RxMC within the framework of the ESPResSo software \cite{weik2019a,weeber24a}, which permits to combine it with the other features already implemented in ESPResSo. 
In particular, we couple it to Langevin Dynamics allowing us to sample the reaction and configuration space simultaneously, as explained in Section \ref{sec:coupling_MD}. 
The inputs of our algorithm are the equilibrium constant $K$ and the initial concentration of the reactants. 
It samples the equilibrium concentration of each reacting species, which can be used to calculate the degree of conversion and other derived quantities.  

In each trial move, the algorithm proceeds by performing the following steps:
\begin{enumerate}
    \item \textit{Selecting the direction of reaction and the reactive groups involved.} 
    A direction of the reaction (forward or reverse) is selected randomly with equal probability. 
    If the forward direction is selected,
    \begin{subequations}
  \begin{equation}
    \label{eq:AAbonding-forw}
    \ce{A + A -> A-A}
  \end{equation}
  \begin{equation}
    \label{eq:ABbonding-forw}
    \ce{ A + B -> A-B}
  \end{equation}
  \label{eq:forward}
\end{subequations}
    a free functional group A is randomly chosen with equal probability.
    Depending on the type of reaction, (Eq. \ref{eq:AAbonding-forw} or \ref{eq:ABbonding-forw}) a second free functional group A or B is randomly chosen with equal probability. 
    If the reverse direction is selected,
    \begin{subequations}
  \begin{equation}
    \label{eq:AAbonding-rev}
    \ce{A-A -> A + A}
  \end{equation}
  \begin{equation}
    \label{eq:ABbonding-rev}
    \ce{A-B  -> A + B }
  \end{equation}
  \label{eq:reverse}
  \end{subequations}
  a bonded pair \ce{A-A} or \ce{A-B} is randomly chosen with equal probability.
  To enable coupling of the eb-RxMC with Molecular Dynamics schemes, the algorithm can be restricted to only select reactive groups at a distance below a given inclusion radius $\rin$.
  Such restriction must be applied both in the forward (Eq. \ref{eq:forward}) and in the reverse (Eqs. \ref{eq:reverse}) directions of the reaction.
    
    \item \textit{Accepting the trial move} with the probability given by the Reaction ensemble\cite{johnson1994,smith1994} 
    \begin{equation}
    P^{\textrm{acc}} =   \min \left[ 1,  W_{\ce{RxMC}}  \right] = \min \left[ 1, (K V^{-1} N_{\ce{A}}^{-1}/\cref)^\xi \exp{(-\beta \Delta U)} \prod_{i} \frac{N_{i}!}{(N_{i}+\nu_{i}\xi)!} \right]  
        \label{eq:Rx_prob}
    \end{equation}
    where $V$ is the volume of the simulation box,  $\beta = 1/(k_{\ce{B}}T)$ is the inverse of the thermal energy, $N_{\ce{A}}$ is the Avogadro number, $\nu_i$ and $N_i$ respectively are the stoichiometric coefficient and the number of the reactive group $i$, where $i \in \{\ce{A,B,A-A,A-B}\}$ is the set of reactive groups involved in the reaction.
    The value of $\xi$ is $\xi=1$ in the forward direction (Eq. \ref{eq:forward}) and it is  $\xi=-1$ in the reverse direction (Eq. \ref{eq:reverse}).
    $\Delta U$ is the change in the potential energy of the system due to the reaction (Eq. \ref{eq:forward} or Eq. \ref{eq:reverse}).
    We emphasize that $\Delta U$ does not include the potential energy of the newly created or deleted bond.
    The Gibbs free energy of the reaction, which results in the bond formation, $\Delta_{\mathrm{r}}G$, includes the potential energy of the bond, as well as the entropy contributions and changes in the solvation free energies of the reacting species.\cite{mcquarrie1997a,atkins98a}
    Therefore,  the potential energy of the newly created or deleted bond is already included in the value of $K$ (Eq. \ref{eq:DeltaG_K}).
    This also implies that the choice of the bonding potential is arbitrary and should not affect the equilibrium degree of conversion.
    If the selection of reactive particles is constrained within $\rin$ in the previous step, then this bias introduced in the probability of selecting the bond partners needs to be reflected in correcting the acceptance probability in Eq. \ref{eq:Rx_prob}.
    In Section \ref{sec:coupling_MD}, we provide a detailed explanation on how to correct for the introduced bias.

    \item \emph{Changing the system to a new state if the trial move has been accepted.} 
    In the forward direction, we add a bonding potential $U_{\ce{bond}}$ between the two selected groups, representing a chemical bond. 
    In the reverse direction, we remove the bonding potential between the selected pair.
\end{enumerate}

\subsection{\label{sec:coupling_MD}Bias of the eb-RxMC to enable coupling with Langevin Dynamics}
\begin{figure}[H]
\centering
\includegraphics [ width =0.99\textwidth]{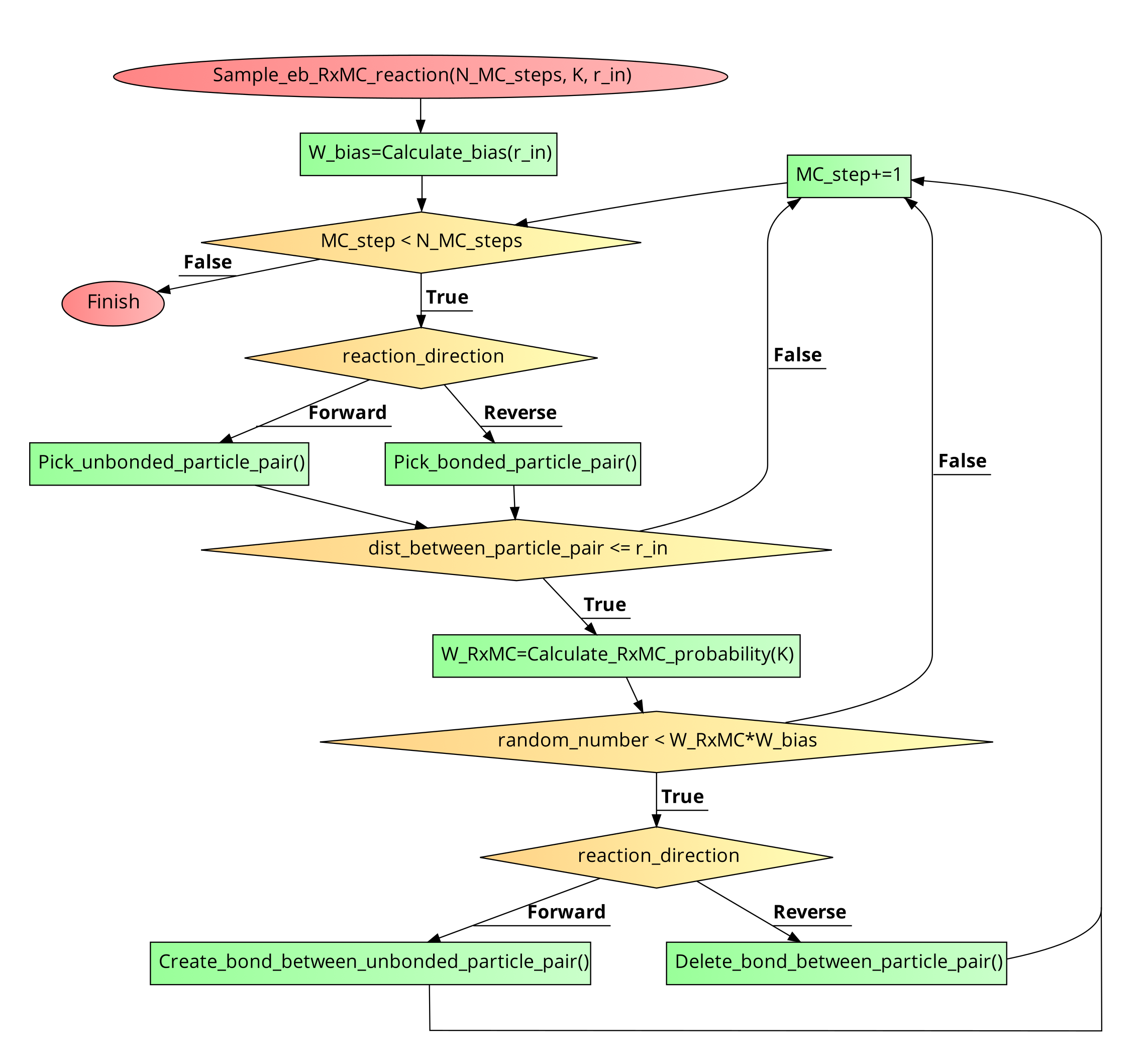}
\caption{Flow diagram of our implementation of the eb-RxMC algorithm. Scheme adapted from a flow diagram originally produced using the software from the website code2flow.com. 
}
\label{fig:code_scheme}
\end{figure}

To enable Langevin Dynamics (LD) integration after a set of eb-RxMC moves, it is necessary to ensure that the forces between the simulated particles are not too big.
Big forces would cause an unstable integration and consequently a failure of the integrator.
To prevent such problems, we biased the eb-RxMC algorithm to only propose creating or deleting bonds between particle pairs separated by a distance below a certain inclusion radius $\rin$, as schematically illustrated in Fig. \ref{fig:code_scheme}.
This radius is an adjustable parameter of the method. 
The specific choice of $\rin$ should not affect the sampling of the equilibrium properties of the system, however, it could affect the sampling efficiency and stability of the subsequent LD.
In general, the value of this radius is chosen as a compromise between a large enough value to enable sufficient sampling while avoiding failure of the LD integrator due to the forces arising when creating new bonds.

To ensure detailed balance, we corrected for the bias in the proposal probability by modifying the acceptance probability accordingly.\cite{frenkel96a}
The sufficient condition to satisfy the detailed balance is that the flow of configurations from an arbitrary state $s_i$ to any other state $s_j$ is equal to flow in the opposite direction,
\begin{equation}
    P_{\ce{obs}}(s_i)P_{\ce{trial}}(s_i \rightarrow s_j)P_{\ce{acc}}(s_i \rightarrow s_j) = P_{\ce{obs}}(s_j)P_{\ce{trial}}(s_j \rightarrow s_i)P_{\ce{acc}}(s_j \rightarrow s_i)
\end{equation}
where $P_{\ce{obs}}(s_i)$ is the probability of observing state $s_i$, $P_{\ce{trial}}(s_i \rightarrow s_j)$ is the probability of proposing a trial move from $s_i$ to $s_j$ and $P_{\ce{acc}}(s_i \rightarrow s_j)$ is the probability of accepting such trial move. 
In absence of any inclusion radius, $P_{\ce{trial}}(s_i \rightarrow s_j)=P_{\ce{trial}}(s_j \rightarrow s_i)$ and  $P_{\ce{acc}}$ is given by Eq. \ref{eq:Rx_prob}. 

The biased proposal probability is different in the forward and reverse direction of the reaction because the probability of finding a pair of \emph{unbonded} particles within $r_{\ce{in}}$ is different from that of finding a pair of \emph{bonded} particles within the same distance, $P_{\ce{trial}}(s_i \rightarrow s_j) \ne P_{\ce{trial}}(s_j \rightarrow s_i)$.
In order to correct for such asymmetric sampling of the reaction space, $P_{\ce{acc}}$ needs to be chosen so that it fulfils the detailed balance condition
\begin{equation}
    \frac{P_{\ce{acc}}(s_i \rightarrow s_j)}{P_{\ce{acc}}(s_j \rightarrow s_i)} = \frac{P_{\ce{trial}}(s_j \rightarrow s_i)}{P_{\ce{trial}}(s_i \rightarrow s_j)} \frac{P_{\ce{obs}}(s_j)}{P_{\ce{obs}}(s_i)}.
    \label{eq:detailed_balance}
\end{equation}
Our choice of $P_{\ce{acc}}$ that satisfies Eq. \ref{eq:detailed_balance} is given by
\begin{equation}
    P_{\ce{acc}} 
    =  \min \left[ 1, 
        \left(\frac{\Pb(\rin)}{\Pu(\rin)} K V^{-1} N_{\ce{A}}^{-1}\right)^\xi 
        \exp{(-\beta \Delta U)} 
        \prod_{\ce{i}} \frac{N_{\ce{i}}!}{(N_{\ce{i}}+\nu_{\ce{i}}\xi)!} 
        \right]
    =\min \left[ 1, W_{\ce{bias}} W_{\ce{RxMC}}  \right].
\label{eq:prob_bias}
\end{equation}
where $\Pu(\rin)$ is the probability of finding two unbonded particles within $\rin$, $\Pb(\rin)$ is the probability of finding two bonded particles within $\rin$ and $W_{\ce{bias}}=(\Pb(\rin)/\Pu(\rin))^\xi$.
In the ideal case, when simulating a system without other interactions than the bonds between the particles, $\Pb(\rin)$ and $\Pu(\rin)$ can be calculated in advance.
On the contrary, when simulating an interacting system, both $\Pb$ and $\Pu$ are affected by other interactions.
In such case, $\Pb$ and $\Pu$ need to be calculated on the fly during the simulation to ensure self-consistency of the sampling.
It follows from Eq.~\ref{eq:prob_bias}  that the restriction of only proposing trial moves between particles at a distance below the inclusion radius must be applied in both directions of the reaction.
More details about the derivation of Eq. \ref{eq:prob_bias} can be found in the Supporting Information (Section S2), including the analytical solutions of $\Pu$ and  $\Pb$ under the ideal gas approximation. 

\subsection{Simulation model\label{sec:model}}
In our simulations, each coarse-grained particle represents one functional group.
The number of particles $\Npart$ remained constant throughout whole simulation but the bonds between them were randomly created or deleted, following the eb-RxMC procedure described in Section \ref{sec:RxMC}. 
We note that the size of the simulation box and concentration can be chosen arbitrarily for an ideal system because the only relevant parameter that determines the equilibrium properties of the system is $\plambda = \pK + \pC$.
Therefore, one can equivalently choose to fix the concentration and vary the value of $K$ or fix the value of $K$ and vary the concentration.
Our choice was to fix the concentration and vary the equilibrium constant of the reaction $K$ as the input of our simulations.
For a non-ideal system, this choice is no longer arbitrary, because the effect of interactions on the equilibrium properties depends on distances between the particles, and hence on their concentration.
The outputs, such as the equilibrium degree of conversion or the chain length distribution in the polycondensation reactions, were then calculated as ensemble averages over the generated configurations, as detailed in Section~\ref{sec:sim_prot}.

The bonds between particles were represented by a harmonic potential
\begin{equation}
    U_{\ce{bond}} = \frac{1}{2}\kh(l-l_0)^2
    \label{eq:harmonic}
\end{equation}
where $l$ is the distance between the bonded particles, $l_0$ is the equilibrium bond length and $\kh$ is the harmonic constant that determines the stiffness of the potential.
This harmonic potential was added and removed between particles following the eb-RxMC procedure described in Section \ref{sec:RxMC}.
Unless otherwise stated, all bonds in our simulation box were harmonic bonds (Eq. \ref{eq:harmonic}) with $\kh = 790$ $\kT/\textrm{nm}^2$ and $l_0 = \qty{0.355}{nm}$.
Similar to the choice of concentration, also the choices of $\kh$ and $l_0$ are arbitrary, as long as we are simulating an ideal system.
Nevertheless, for future reference, we set them to values commonly used in coarse-grained models of polymers.\cite{lunkad21a,lunkad22a}

For the dimerization reactions between identical monomers, Eq. \ref{eq:AAbonding}, our initial configuration contained $\Npart$ identical particles of type A.
Then, we varied the equilibrium constant as the input parameter of the simulation.
For the dimerization reactions between different monomers, Eq. \ref{eq:ABbonding}, our initial configuration contained $f_\mathrm{A} \Npart$ particles of type A and $(1-f_\mathrm{A})\Npart$ particles of type B.
Then, in addition to the variation of equilibrium constant, we also varied the initial composition of the reacting mixture by varying the value of $f_\mathrm{A}$.
Our simulation box size is $L = 30 l_0 = \qty{10.65}{nm}$. 
With $\Npart =50$ particles in the box, the molar concentration of monomers becomes $\ctot = \qty{70}{mM}$.

For the polycondensation reactions, \ref{eq:l_polymerization}, we represented each monomer by two particles of the same type, corresponding to two reactive functional groups.
These two particles were connected by a permanent bond, which was not considered in the eb-RxMC procedure.
Furthermore, we prevented the formation of rings by not permitting intra-molecular reactions.
In this case, we used a larger box size $L = 50 l_0 = \qty{17.75}{nm}$ and larger number of particles $\Npart = 200$ particles, yielding a molar concentration of monomers of $\ctot = \qty{60}{mM}$.
Otherwise, all parameters used in the simulations of polycondensations were the same as for dimerizations.

\subsection{Simulation protocol\label{sec:sim_prot}}

We used the ESPResSo v4.2.1.\cite{weik2019a, weeber24a} simulation software to perform the simulations.
One simulation cycle consisted of trial reaction moves equal to the number of particles $N_{\textrm{trial}}=\Npart$, followed by $N_\mathrm{MD}=4\cdot10^{4}$ integration steps of the Langevin dynamics.
A typical simulation consisted of 30000 such cycles, yielding a total  $t_{\mathrm{sim}} = 1.2\cdot10^7\tau$ of Langevin dynamics time evolution and $6\cdot10^6$ reaction trial moves, consuming  12 hours of computing time on the MetaCentrum computing cluster, typically using nodes with Intel(R) Xeon(R) Gold 6130 CPU @ 2.10GHz, 192 GB RAM and 1x 960 GB NVMe of disk.
Unless otherwise stated, the reactions were sampled using the eb-RxMC algorithm described in Section \ref{sec:RxMC} with an inclusion radius of $\rin=\qty{3.8}{nm}$.

The LD simulations were performed using a time step $\delta t = 0.01\tau$, and a damping constant $\gamma = 1.0/\tau$, where $\tau = l_0 \sqrt{m/\kT}$ with a reduced temperature $T = \kh l_0^2 / \kb$.
Similar to the choice of $\kh$, also the choices of temperature $T$ and particle mass $m$ are arbitrary and have no effect on the computed equilibrium properties, as long as we are simulating an ideal system.
During the simulation, we were estimating on the fly the probability density of bond lengths, $P(l)$.
All sampled quantities were stored with time intervals $t_{\mathrm{coord}} = \qty{400}{\tau}$ and were averaged after discarding the first 30\% of each simulation run as equilibration, corresponding to the first $3.6\cdot10^6\tau$ of the simulation. 
Statistical uncertainty of the computed averages was estimated using the block analysis method.\cite{janke2002a}

\section{Results and discussion\label{sec:results}}
We aim to  validate our implementation of the eb-RxMC algorithm using simple systems, for which exact analytical solutions are available.
Therefore, we restrict the results in this article to ideal non-interacting systems, in which the only interaction present is the  harmonic bond potential (Eq. \ref{eq:harmonic}), representing both reversible and permanent bonds between particles.
First, we use dimerization reactions to show that our eb-RxMC simulations yield the correct degree of conversion and bond length distribution of the reversible bonds, regardless of the choice of the equilibrium constant $K$, stiffness of the harmonic bond $k_\mathrm{h}$ or inclusion radius $r_\mathrm{in}$ for the biased sampling.
Next, we simulate linear polycondensation reactions to prove that our eb-RxMC algorithm also samples the correct equilibrium distribution of polymer chain lengths, quantified by the mass fraction $w(M)$, and end-to-end distances, $R_{\mathrm{e}}(M)$, for each chain length $M$.

\subsection{Dimerization reactions \label{sec:dimerization} }

\begin{figure}[H]
\centering
\includegraphics [ width =0.49\textwidth]{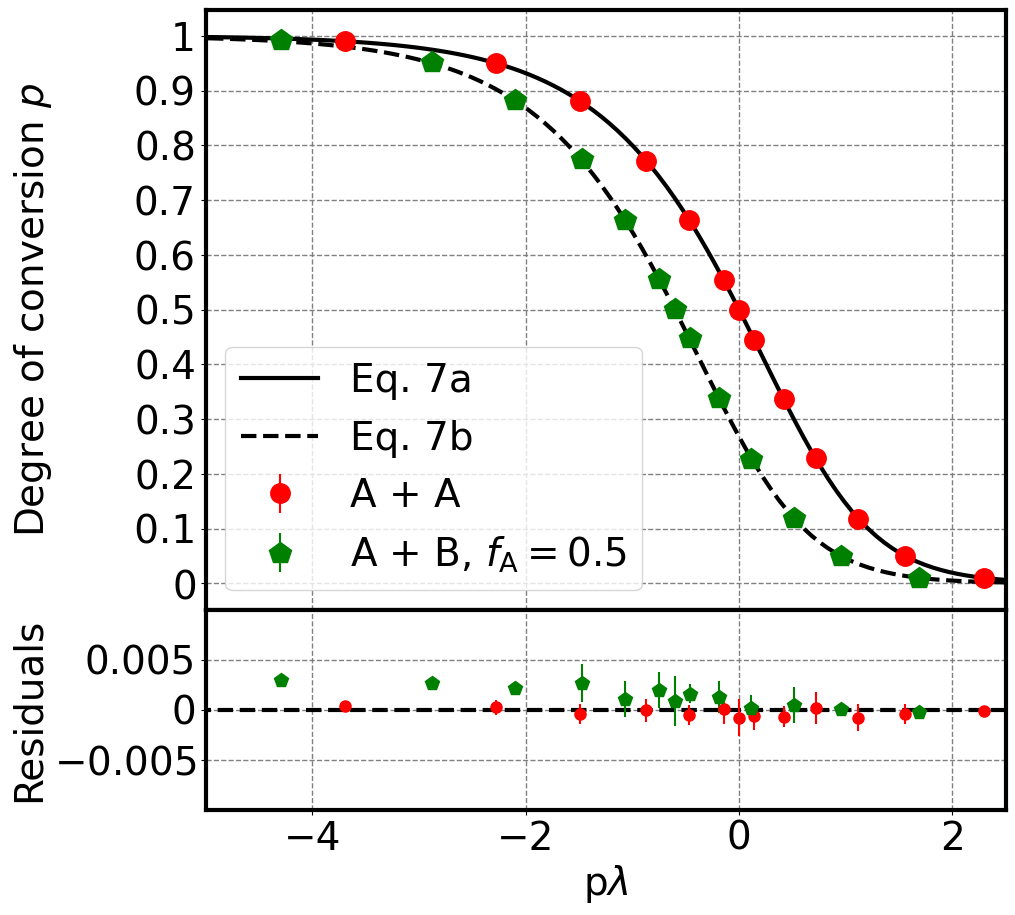}
\includegraphics [ width =0.49\textwidth]{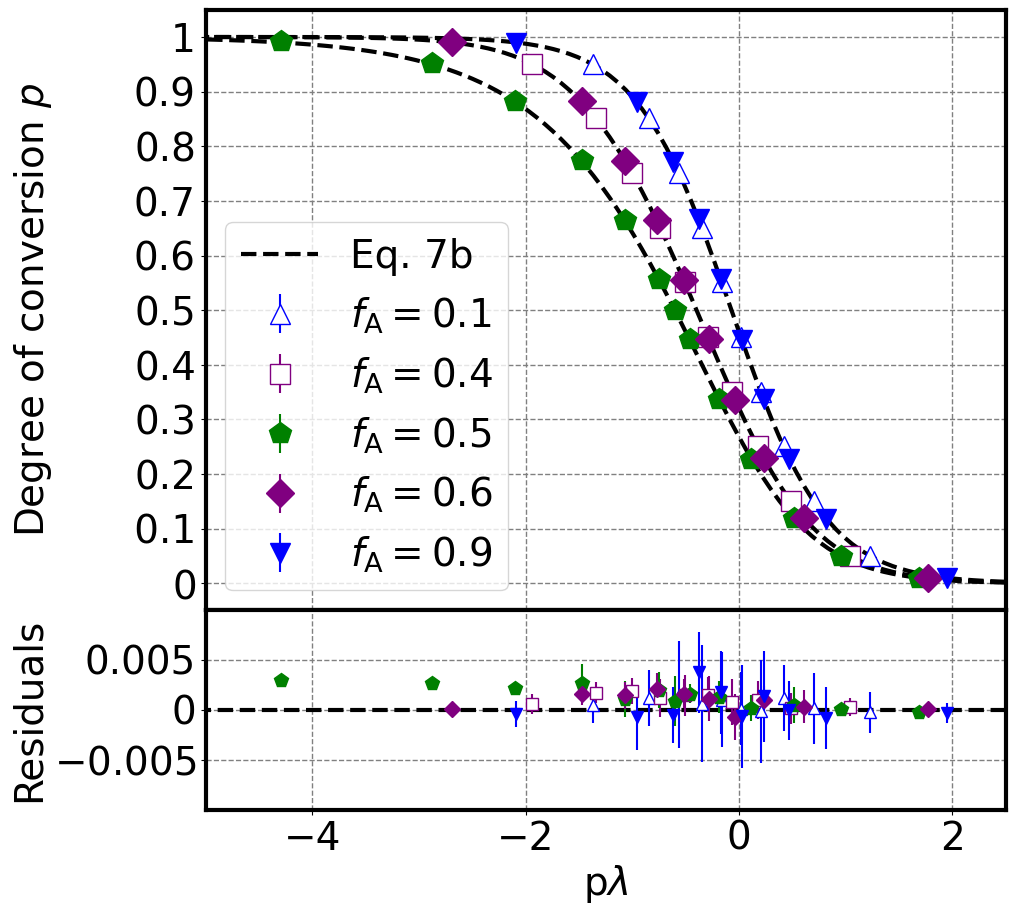}
\caption{
Left panel: Degree of conversion $p$ as a function of $\plambda$ (Eq. \ref{eq:plambda}) for A+A dimerization reactions (Eq. \ref{eq:AAbonding}) and A+B dimerization reactions (Eq. \ref{eq:ABbonding}) at equimolar composition.
Right panel: $p$ as a function of $\plambda$ for A+B dimerization reactions at various initial compositions, defined by the mole fraction of A $f_\mathrm{A}$.
The eb-RxMC simulations (markers) are compared with analytical solutions given by Eq. \ref{eq:p_AA_id} (continuous line) and \ref{eq:p_AB_id} (dashed lines). 
 Residuals (bottom panels) are calculated as the difference between the simulated $p$-value and the theoretical $p$-value (Eqs. \ref{eq:p_AA_id}-\ref{eq:p_AB_id}).
   }
\label{Fig:p_id}
\end{figure}

We start with dimerization reactions,  the same test case used by Smith and Triska to validate the original RxMC algorithm.\cite{smith1994} 
To independently validate the sampling of the reaction space, we did not sample the configuration space in these simulations, therefore we did not use any inclusion radius $\rin$.  
In Fig. \ref{Fig:p_id} (left panel), we present the degree of conversion $p$ measured in our eb-RxMC simulations (markers) as a function of $\plambda = \pK + \pC$, where $K$ is the equilibrium constant and $C$ is the relative concentration of the reacting species, as defined in Eq. \ref{eq:plambda}.
The dimerization between two chemically identical monomers (Eq. \ref{eq:AAbonding}) reaches $p=0.5$ at $\plambda = 0$. 
Fig. \ref{Fig:p_id} (left panel) shows that our simulation results reproduce the analytical reference result of Eq. \ref{eq:p_AA_id} with a very high accuracy.
Dimerization between two chemically different monomers, (Eq. \ref{eq:ABbonding}), at equimolar stoichiometric ratio reaches $p=0.5$ at a slightly lower value of $\plambda$ but otherwise the dependence $p(\plambda)$ results in curve with a very similar shape.
The residuals in Fig. \ref{Fig:p_id} are calculated as the difference between the simulated $p$-value and the exact $p$-value from the theory.
They show that the agreement between the numerical results and the analytical theory is excellent.
We only observe a very small systematic deviation in some corner cases with a high degree of of conversion, which are presumably related to finite-size effects, as discussed in Ref.\cite{hebbeker2023}
For dimerization between two chemically different species, we further investigate how the degree of conversion is affected by the initial composition.
In Fig. \ref{Fig:p_id} (right panel) we show simulation results at various initial compositions of the system, ranging from $f_\mathrm{A}=0.1$ to $f_\mathrm{A} = 0.9$.
Once again, we observe a very good agreement between our simulations and analytical results for the given composition.
When either of the reactants is in excess, the $p$-values of A+B reactions as a function of $\plambda$ are increased, as compared to $f_\mathrm{A}=0.5$.
Interestingly, the curves become almost symmetric around $\plambda=0$ at very high or very low value of $f_\mathrm{A}$.
Based on the above comparison, we conclude that our eb-RxMC simulations correctly sample the reaction space for a broad range of equilibrium constants $K$ and initial compositions of the system, in absence of any inclusion radius $\rin$.

\begin{figure}[H]
\centering
\includegraphics [ width =0.49\textwidth]{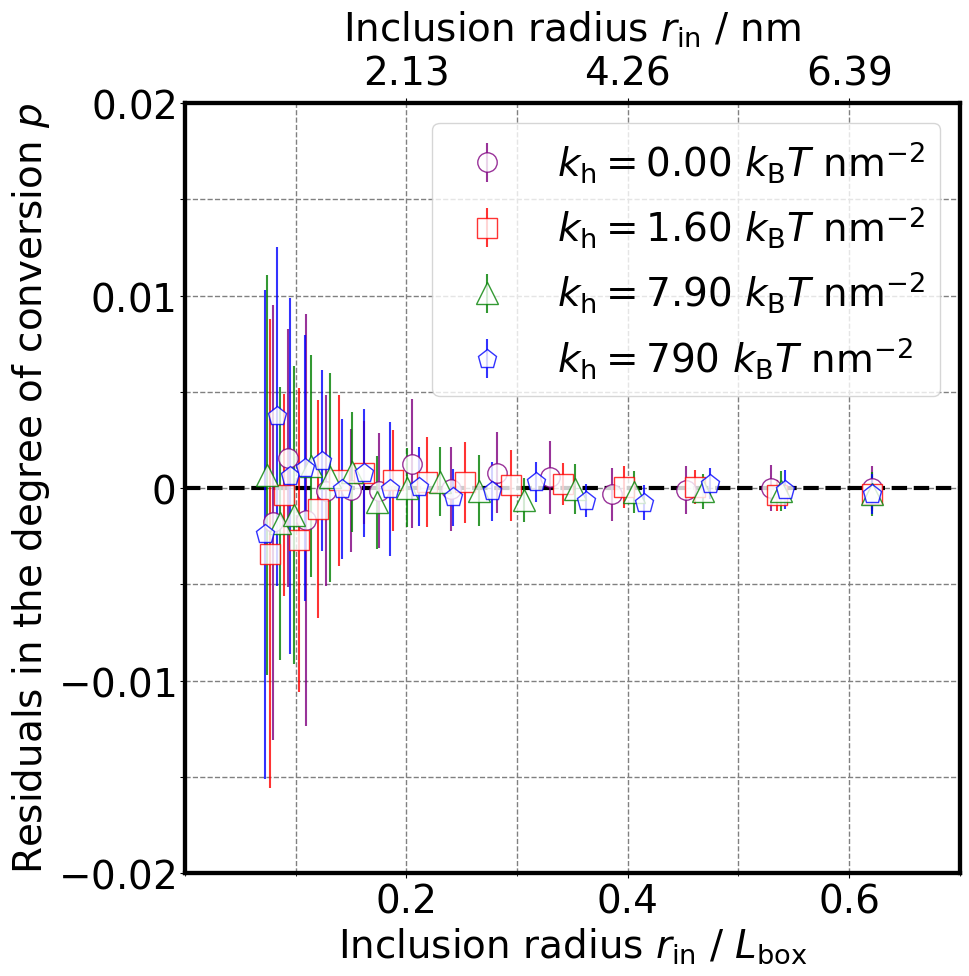}
\includegraphics [ width =0.49\textwidth]{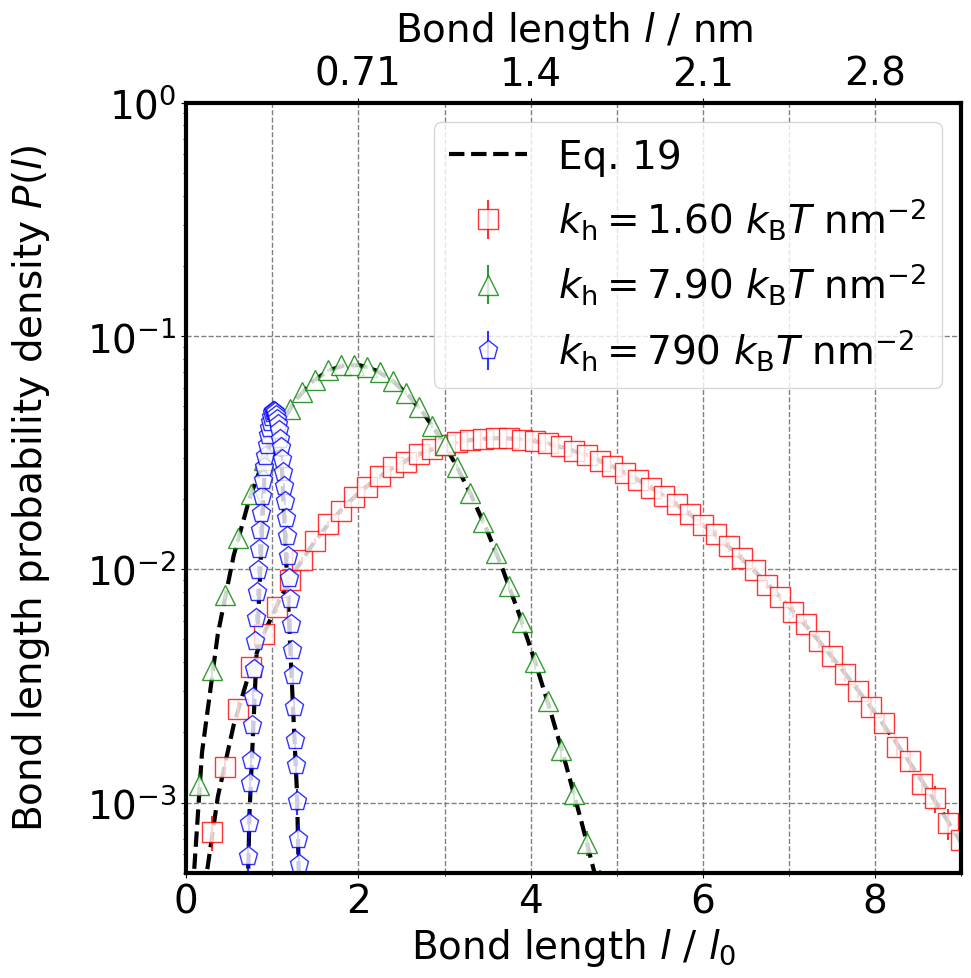}
\caption{
Left panel: Residuals in the degree of conversion $p$ as a function of the inclusion radius $\rin$ at various values of the harmonic constant for the reversible bond $\kh$.
The residuals are calculated as the difference between the measured value in our simulations and the expected value $p=0.8$ given by Eq. \ref{eq:p_AA_id} at $\plambda=-1$.
Right panel: Probability density $P(l)$ of observing a reversible bond with a length $l$ in a bin at a distance $r$ with size $\mathrm{d}r$ at different values of $\kh$.  
The numerical results from our simulations (markers) follow the exact result given by Eq. \ref{eq:P_bond_dis} (dashed lines).}
\label{Fig:r_inc}
\end{figure}

Next, we validate that our eb-RxMC algorithm can be combined with Langevin Dynamics (LD) to simultaneously sample the reaction and configuration space without introducing any artifacts.
When combining both simulation schemes, we use the biased eb-RxMC, sampling the reactions only within an inclusion radius $r_{\ce{in}}$.
We aim to verify that the inclusion radius can be selected almost arbitrarily. 
The only practical constraint is that the forces due to the harmonic potential within this range are small enough to ensure that the subsequent LD integration remains stable.
Furthermore, we aim to verify that the degree of conversion is not affected by the arbitrary choice of the stiffness constant  of the harmonic potential (Eq. \ref{eq:harmonic}).
For this purpose, we consider the case of a A+A dimerization with $\plambda = -1$, corresponding to an expected relatively high  degree of conversion $p=0.8$ (Eq. \ref{eq:p_AA_id}).
This $\plambda$-value is a convenient choice for the later study of polycondensation reactions, because it 
permits to measure a relatively wide distribution of polymer chains of different lengths.
In Fig. \ref{Fig:r_inc} (left panel), we show the residuals in the degree of conversion $p$ as a function of the chosen value of $\rin$ at various values of $\kh$. 
All these residuals are very close to zero, and appear randomly distributed around this reference line, confirming that both arbitrary choices, $\rin$ and $\kh$, have no effect on the computed degree of conversion and that our combination of eb-RxMC with LD yields the correct results.

The choice of $\rin$ only affects the efficiency of the eb-RxMC sampling but does not change the ensemble averages of any observable.
Therefore, one can choose $\rin$ completely arbitrarily, as long as a pure MC simulation scheme is used.
The only criterion for the optimum value of $\rin$ is the sampling efficiency.
When simulating interacting systems using a combination of eb-RxMC and LD, the optimum value of $\rin$ should be chosen as a compromise between higher sampling efficiency of the eb-RxMC algorithm (high $\rin$ values) and a stable integration of the chosen MD scheme (low $\rin$ values).
Our preliminary eb-RxMC simulations with systems including Lennard-Jones and electrostatic interactions suggest that values of $\rin \approx 1$ nm are a good compromise between a sufficient sampling  and a stable integration of LD.
However, this value of $\rin$ should be taken only as a reasonable choice and, in general, one should tune $\rin$ for each specific system of study. 

To prove that our simulations also correctly sample the configuration space, we measured the equilibrium distribution of bond lengths of the reversible bonds at different values of $k_{\ce{h}}$. 
For a harmonic oscillator, this equilibrium distribution can be calculated exactly as the probability density
\begin{equation}
    P(l)  = \frac{l^2 \mathrm{e}^{-\beta U_{\ce{bond}}(l)}}{\int_0^\infty r^2 \mathrm{e}^{-\beta U_{\ce{bond}}(r)} \ce{d}r}.
    \label{eq:P_bond_dis}
\end{equation}
For the harmonic potential, $P(l)$ depends on the chosen value of $k_{\ce{h}}$ \textit{via} $U_{\ce{bond}}$ (Eq. \ref{eq:harmonic}).
Fig. \ref{Fig:r_inc} (right panel) proves that our simulations perfectly match the theoretical prediction given by Eq. \ref{eq:P_bond_dis}.
This result confirms that our simulations properly sample not only the reaction space but also the configuration space of the dimerization reactions.
For completion, we provide an analogous analysis for the case of A+B dimerization at equimolar conditions ($f_\mathrm{A}=0.5$) in the Supporting Information (Section S3). 
Our results show that our combined scheme LD/eb-RxMC also samples correctly the reaction and the configuration space in this case with no significant deviations from the reference analytical equations for the ideal case.

\subsection{Polycondensation reactions\label{sec:polycondensation}}
 
After validating our eb-RxMC implementation for dimerization reactions, we would like to apply it to a more complex problem.
Ideal polycondensation reactions are a suitable complex problem, thanks to the analytical results for chain length distribution\cite{flory36a} and for end-to-end distance.\cite{Flory1969,rubinstein03a}
In this section, we show that our eb-RxMC implementation quantitatively reproduces these analytical results.

\begin{figure}[H]
\centering
\includegraphics [ width =0.99\textwidth]{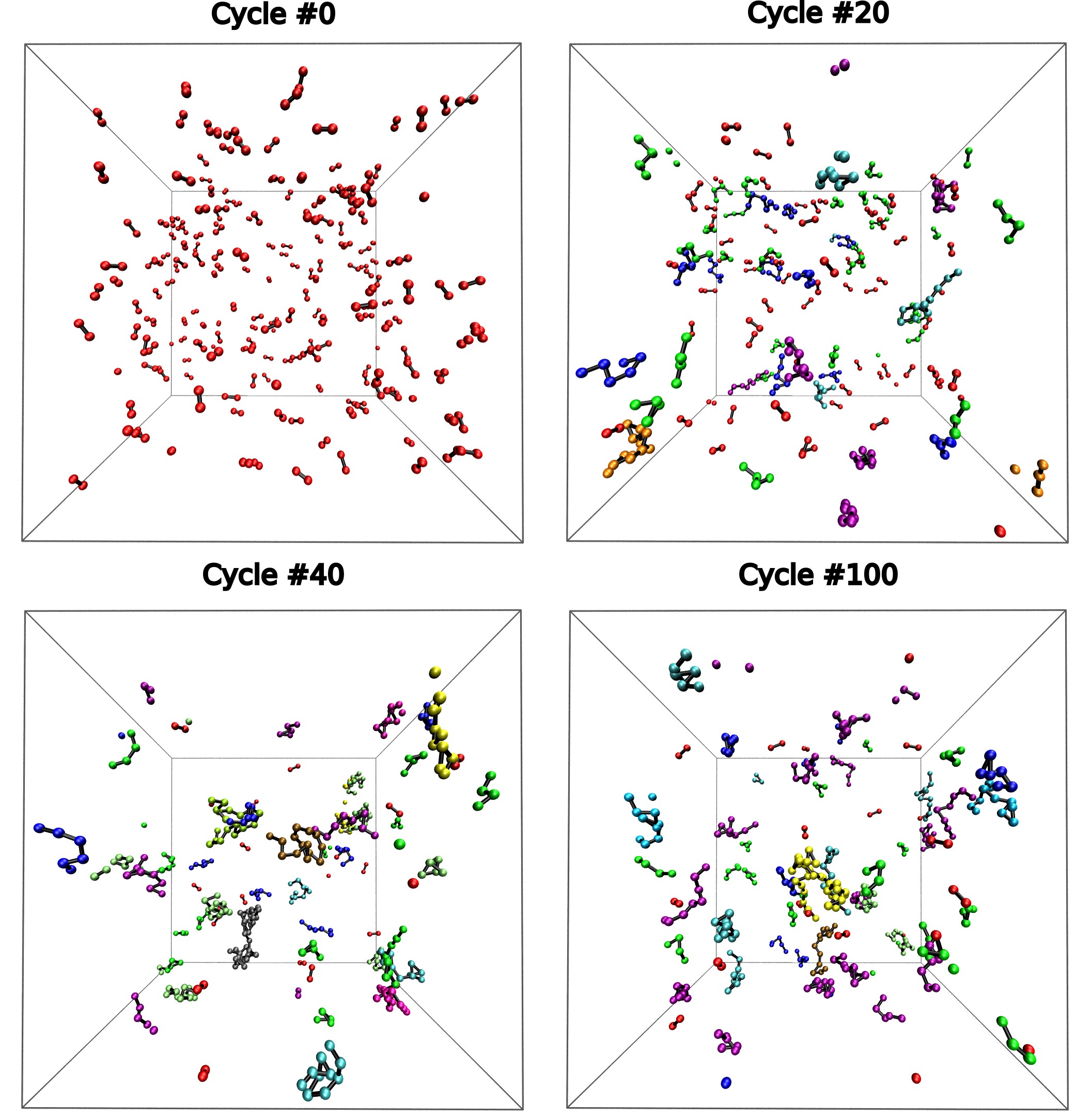}
\caption{Simulation snapshots of different configurations obtained with our eb-RxMC algorithm with explicit reversible bonds. The simulation is initialized with a system consisting in monomers, each containing two reactive particles (Cycle \#0). Throughout the simulation, these monomers can reversibly undergo linear polycondensation reactions (Eq. \ref{eq:l_polymerization}) yielding a distribution of chains with different lengths. For clarity, polymer chains with the same number of monomers have been colored with the same color.
}
\label{Fig:snapshots}
\end{figure}

For illustration, we start with $\plambda = -1$, which yields an expected relatively high  degree of conversion $p=0.8$ (Eq. \ref{eq:p_AA_id}) and a broad distribution of chain lengths.
In Fig. \ref{Fig:snapshots}, we show snapshots of the simulation of a polycondensation reaction at different simulation cycles.
In this figure, each colour represents chains of a specific chain length $M$. 
In cycle \#0, the starting configuration contains only monomers,  {\ie} $p=0$.
As the simulation progresses, the particles reversibly bind and unbind to each other, causing that longer chains gradually appear, until an equilibrium state is reached.
In equilibrium, the number of chains and their length both fluctuate during the simulation, although the total number of particles in solution remains constant.   
  
\begin{figure}[H]
\centering
\includegraphics [ width =0.49\textwidth]{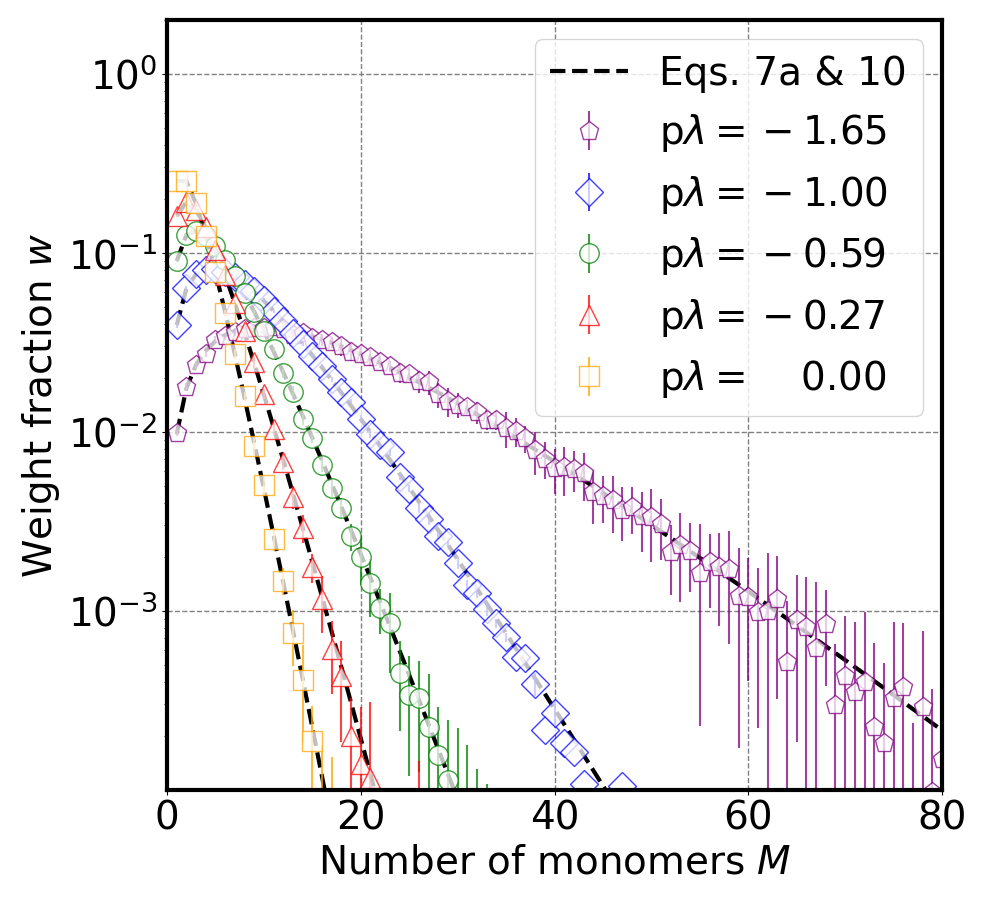}
\includegraphics [ width =0.49\textwidth]
{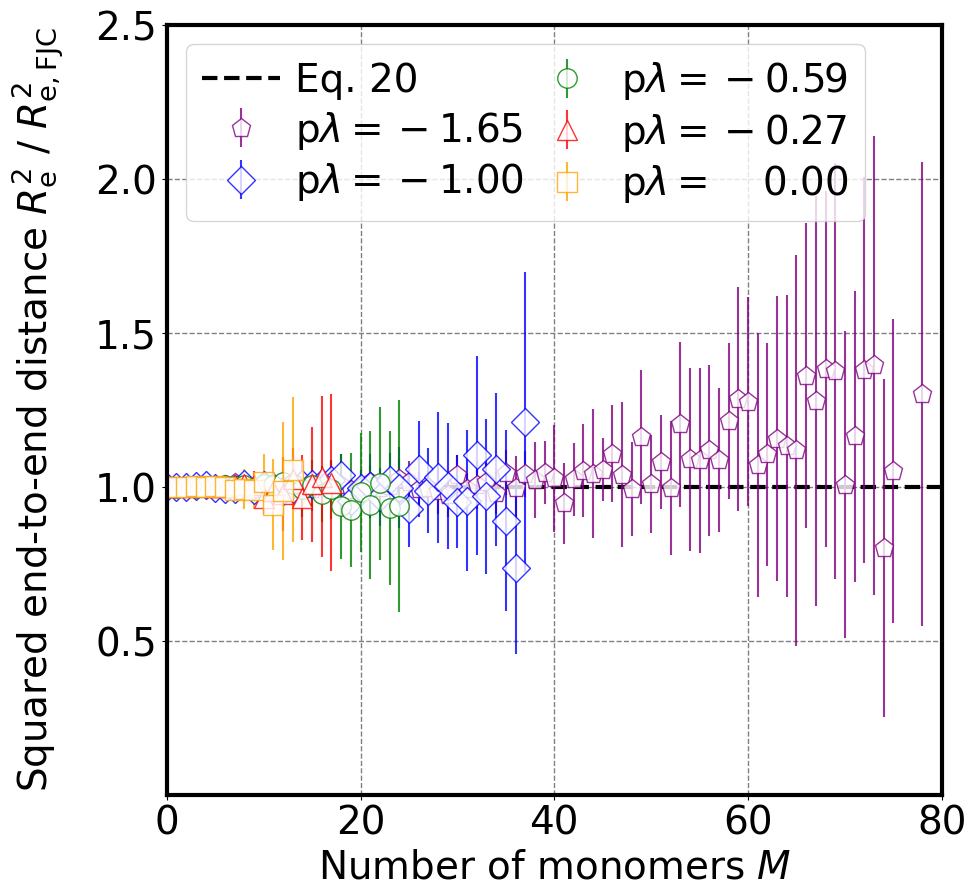}
\caption{Left panel: mass fraction $w$ as a function of the number of monomers in the polymer chain $M$ at various  $\plambda$ values. Markers follow the data measured in our eb-RxMC simulation and the dashed line follows the exact value given by Eqs. \ref{eq:p_AA_id} and \ref{eq:Flory}. Right panel: end-to-end distance $R_{\mathrm{e}}$ as a function of $M$ at various $\plambda$ values (markers). Markers follow the data measured in our simulation and the dashed line follows the exact value given by Eq. \ref{eq:r2_FJC}. For convenience, all results are normalized by $R_{\mathrm{e,\mathrm{FJC}}}^{2}$.}
\label{fig:polycond_panels}
\end{figure}

In Fig. \ref{fig:polycond_panels} (left panel), we compare the simulated mass fraction $w(M)$ with the analytical result given by Eq. \ref{eq:Flory} for various values of $\plambda$, corresponding to various degrees of conversion.
We observe a very good agreement for all values of $\plambda$, even for rather long chains with mass fractions below 0.001.
We note, however, that for long chains the $w(M)$ distributions obtained from simulations can deviate from the analytical results due to finite-size effects,\cite{hebbeker2023} as we demonstrate in the Supporting Information (S4).
Therefore, it is important to ensure that the expected average chain length is much lower than the number of monomers in the simulation box.
As a rule of thumb, the total number of monomers should be at least 10 times higher than the average chain length.

To validate the sampling of configurational space, in Fig. \ref{fig:polycond_panels} (right panel), we plot the simulated end-to-end distance of the chains, $R_{\mathrm{e}} (M)$, divided by the exact analytical results  for the freely jointed chain (FJC)\cite{Flory1969,rubinstein03a}
\begin{equation}
    R_{\mathrm{e,\mathrm{FJC}}}^{2}(M) = b^2 \, (2M-1)
    \label{eq:r2_FJC}
\end{equation}
where $b$ is the Kuhn segment length of the FJC.
In this comparison, we assume that the Kuhn segment length of the FJC can be approximated by the average bond length measured in the simulation, $b \approx \left< l \right>\gtrsim l_0$.
For $R_{\mathrm{e}}$, our simulation results once again match the theory within statistical error.
The simulation results deviate from Eq. \ref{eq:r2_FJC} for the longer chains of each distribution, but these chains occur only rarely in our simulations, which is also reflected in the high statistical uncertainty of their estimated end-to-end distance.
Altogether, our validation for polycondensation reactions has shown that our simulations correctly sample both the reaction and configuration space of polymer chains formed during the reaction.

\section{Conclusions}
We have designed a biased Monte Carlo algorithm to sample chemical equilibrium in the Reaction ensemble by creating explicit bonds between particles in the simulation box.
We refer to such algorithm as explicit bonding Reaction Ensemble Monte Carlo (eb-RxMC) to distinguish it from the original implementation of the reaction ensemble, in which the reaction is represented by inserting the newly formed particles and deleting the ones which have reacted.
The eb-RxMC algorithm will become publicly available as a module in one of the future releases of ESPResSo.\cite{weik2019a,weeber24a}. 

Our eb-RxMC algorithm is biased to sample only the reaction within an inclusion radius $\rin$.
This bias allows us to couple the eb-RxMC with Langevin Dynamics to sample the configuration space.
The value of $\rin$ can be chosen arbitrarily but, in order to enable this coupling, it should be chosen such that big forces do not occur due to newly created bonds, so that the LD integration remains stable.
By simulating systems for which exact analytical solutions are known, we demonstrated that our algorithm correctly samples both the reaction and configuration space.
Namely, we showed that for dimerization reactions with different equilibrium constants and initial compositions, the degree of conversion measured in our simulations perfectly matched the reference value given by the analytical equations, irrespective of the choice of the inclusion radius $\rin$ or the stiffness of the harmonic potential, $\kh$.
Next, we showed that our simulations can correctly match the analytical chain length distribution\cite{Flory1936} and end-to-end distance\cite{Flory1969,rubinstein03a} of ideal chains in polycondensation reactions.
Therefore, we conclude that our eb-RxMC simulations correctly sample both the reaction and the configuration space of these reference systems. 

In this work, we have validated our algorithm for a set of ideally behaving systems, for which analytical results are available.
The interesting applications of our algorithm consist in using it for interacting systems, for which analytical results are not available, and therefore, their behaviour is difficult to predict from theory.
However, such applications would not be trustworthy without a previous validation.
For example, we are currently using our eb-RxMC algorithm to simulate polymers containing derivatives of boronic acids in the side chains.\cite{Vdorvdovivc2019,Markova2021}
In these polymers, the formation of hydrogen bonds competes with the acid-base ionization equilibrium.
Both the hydrogen bonds and ionization of the side-chains affect the chain conformation, creating a complicated feedback loop, which is difficult to capture by intuition.
Another example includes hydrogen bonding in linear poly(ethyleneimine), which has been proposed as a hypothetical explanation for the peculiar features of titration curve of this molecule.\cite{Garces2014}
Our preliminary results suggest that our algorithm captures well the competition of the hydrogen bonding with the acid-base equilibrium in these systems, being able to explain features of the potentiometric titration curves that cannot be justified by purely electrostatic arguments. 
In a broader scope, our algorithm has potential applications in various other dynamical macromolecular systems, in which the formation of reversible bonds is one of the governing factors.

\section{Supplementary Material}
Additional material provided in the Supporting Information: (i) detailed derivation for the degree of conversion reactions in the ideal gas limit (Eqs. \ref{eq:p_AA_id} and \ref{eq:p_AB_id}); (ii) detailed derivation for the acceptance probability of the eb-RxMC method (Eq. \ref{eq:prob_bias}); (iii)  additional plots showing that the eb-RxMC correctly samples the reaction and configuration space of A+B dimerization reactions with $f_\textrm{A}=0.5$; and (iv)  a brief discussion on finite size effects in eb-RxMC simulation of linear polycondensation reactions.

\begin{acknowledgement}
We thank Dr. Rita S. Dias for useful discussions.
We acknowledge the financial support of the Czech Science Foundation, Grant 21-31978J, and the funding from the Norway grants, provided by the Czech Ministry of finance, project number EHP-BFNU-OVNKM-4-215-01-2022.
P.M.B. acknowledges the European Union’s Horizon Europe research and innovation program under the Marie Sklodowska-Curie grant agreement No 101062456 (ModEMUS).
Computational resources were supplied by the project "e-Infrastruktura CZ" (e-INFRA CZ LM2018140) supported by the Ministry of Education, Youth and Sports of the Czech Republic.
\end{acknowledgement}

\bibliography{icp,Bibliography}

\section{Supporting information}

\subsection{Derivation of the degree of conversion for dimerization reactions in the ideal gas limit}

Let us start with the reversible association between two identical functional groups A,
\begin{equation}
\ce{2A <=> A-A}
\label{eq:AAbonding}
\end{equation}
where \ce{A-A} represents a pair of bonded A groups.  
The chemical reaction equilibrium in Eq. \ref{eq:AAbonding} is determined by the corresponding equilibrium constant
\begin{equation}
K \equiv  \frac{ a_{\ce{A-A}} }{a_{\ce{A}}^2},
\label{eq:K_AA}
\end{equation}
where $a_x$ is the activity of the group $x \in\{\ce{A, A-A}\}$. 
In the ideal gas limit, \emph{i.e.} assuming a non-interacting system, the activities in Eq. \ref{eq:K_AA} can be replaced by concentrations $a_{\ce{x}}=c_{\ce{x}}/\cref$, yielding
\begin{equation}
K \overset{\mathrm{ideal}}{=} \frac{c_{\ce{A-A}} \, \cref}{c_{\ce{A}}^2}
\label{eq:K_AA_id}
\end{equation}
where $\cref = 1 \ce{mol/kg}$ is the reference concentration, following the definition by the International Union of Pure and Applied Chemistry (IUPAC).\cite{mcnaught1997} 
The choice of $\cref$ defines the value of the chemical potential and simultaneously ensures that $K$ is dimensionless.
For the reaction in  Eq. \ref{eq:AAbonding}, we define the degree of conversion as the fraction of the concentration of \ce{A-A} pairs over their maximum attainable concentration, $\cmax_{\ce{A-A}}$, 
\begin{equation}
p \equiv \frac{c_{\ce{A-A}}}{\cmax_{\ce{A-A}}} = \frac{ \ctot -c_{\ce{A}}}{\ctot},
\label{eq:p_AA}
\end{equation}
where $\ctot$ is the total concentration of reactive groups, \ie $\ctot=c_{\ce{A}} + 2c_{\ce{A-A}}$.
The concentration of each of the reactive groups can be expressed using the degree of conversion,
\begin{subequations}
\begin{equation}
c_{\ce{A}} = \ctot(1-p)
\,,
\label{eq:c_A}
\end{equation}
\begin{equation}
c_{\ce{A-A}} = \frac{1}{2}\ctot p
\,.
\label{eq:C_AA}
\end{equation}
\end{subequations}
The equilibrium constant can be reformulated in therms of the degree of conversion inserting Eqs. \ref{eq:c_A} and \ref{eq:C_AA} into Eq. \ref{eq:K_AA_id},
\begin{equation}
K \overset{\mathrm{ideal}}{=} \frac{\cref}{2\ctot} \frac{p}{(1-p)^2}.
\label{eq:Kx_AA_id}
\end{equation}
Eq. 7a in the main text is obtained by solving Eq. \ref{eq:Kx_AA_id} for $p$,
\begin{equation}
p = 1 + \frac{1-\sqrt{1+8 \lambda}}{4 \lambda} \, ,
\label{eq:p_AA_id}
\end{equation}
where we have defined $\lambda$ as
\begin{equation}
\plambda = -\log_{10}(\lambda) = \pK + \pC \, , 
\label{eq:lambda}
\end{equation}
or alternatively $\lambda = K \ctot / \cref$, where $\pK=-\log_{10}(K)$ and $\pC = -\log_{10}(\ctot/\cref)$. 
The definition of $\plambda$ simplifies the functional form of Eq. \ref{eq:p_AA_id} and it casts the equation in the logarithmic form, which is convenient when using $\pK$ instead of $K$.

In the case of dimerization reactions between two chemically different functional groups A and B,
\begin{equation}
\ce{A} + \ce{B <=> A-B} \, ,
\label{eq:ABbonding}
\end{equation}
the derivation of the degree of conversion proceeds similar to Eq. \ref{eq:AAbonding}. 
The equilibrium constant for Eq. \ref{eq:ABbonding} is given by
\begin{equation}
K \equiv  \frac{ a_{\ce{A-B}} }{a_{\ce{A}} a_{\ce{B}}}
\,,
\label{eq:K_AB}
\end{equation}
where $a_{\ce{x}}$ is the activity of the chemical species $x \in\{ \ce{A, B, A-B}\}$. 
In the ideal gas limit, $K$ can be expressed in terms of the concentrations of reactants
\begin{equation}
K \overset{\mathrm{ideal}}{=} \frac{c_{\ce{A-B}} \, \cref}{c_{\ce{A}}c_{\ce{B}}}
\,.
\label{eq:K_AB_id}
\end{equation}
Following analogous arguments as than in the case of dimerization reactions (Eq. \ref{eq:p_AA}), we define the degree of conversion for Eq. \ref{eq:ABbonding} as the fraction of the concentration of \ce{A-B} pairs over their maximum attainable concentration, $\cmax_{\ce{A-B}}$, 
\begin{equation}
p \equiv \frac{c_{\ce{A-B}}}{\cmax_{\ce{A-B}}} = \frac{\ctot -c_{\ce{A}} -c_{\ce{B}}}{\ctot (1-\rex)},
\label{eq:p_AB}
\end{equation}
where $\ctot$ is the total concentration of reactive groups, \ie $\ctot=c_{\ce{A}} + c_{\ce{B}} + 2c_{\ce{A-B}}$ and $\rex$ is the excess ratio, defined as
\begin{equation}
\rex \equiv \frac{c_{\ce{ex}}}{\ctot} = | 2f_{\ce{A}} -1 |
\label{eq:rex}
\end{equation}
where $c_{\ce{ex}} = | c_{\ce{A}} - c_{\ce{B}} |$ is the excess concentration of reactive groups and $f_{\ce{A}}$ is the mole fraction of A groups in the reactive mixture, including both bonded and free A groups. 
By introducing $\rex$ in Eq. \ref{eq:p_AB}, we ensure that $p=1$ corresponds to the maximum attainable conversion of the reaction, independently of the initial composition of the reacting system.

From now on, we will assume that reactant B is in excess.
This comes without loss of generality because A and B had equivalent roles in the preceding equations.
Then concentration of individual reactants can be expressed using $p$ and $\rex$ as
\begin{subequations}
\begin{equation}
c_{\ce{A}} = \frac{1}{2} \ctot (1-p) (1-\rex) = \frac{1}{2} \ctot (1-p - \rex + p\rex)
\,,
\label{eq:c_A'}
\end{equation}
\begin{equation}
c_{\ce{B}} = \frac{1}{2} \ctot (1-p) (1-\rex) + \ctot \rex = \frac{1}{2} \ctot (1-p+\rex+p\rex)
\,,
\label{eq:c_B'}
\end{equation}
\begin{equation}
c_{\ce{A-B}} = \frac{1}{2}\ctot p (1-\rex)
\,.
\label{eq:C_AB'}
\end{equation}
\end{subequations}
Inserting Eqs. \ref{eq:c_A'}-\ref{eq:C_AB'} into Eq. \ref{eq:K_AB_id}, $K$ can be expressed in therms of $p$ and $\rex$,
\begin{equation}
K \equiv  \frac{2\cref}{\ctot} \frac{p}{(1-p)(1-p+\rex+p\rex)}.
\label{eq:Kx_AB_id}
\end{equation}
Finally, Eq. 7b in the main text is obtained by solving Eq. \ref{eq:Kx_AB_id} for p,
  \begin{equation}
    p =  (1-r_{\ce{ex}})^{-1} \left(1+ \frac{1-\sqrt{1 + 2\lambda+\lambda^2r_{\ce{ex}}^2}}{\lambda}\right) \,, 
    \label{eq:p_AB_id}
  \end{equation}
where $\plambda$ is given by Eq. \ref{eq:lambda}.

\subsection{Derivation of the acceptance probability for the explicit bonding Reaction ensemble Monte Carlo}

Below, we show that our algorithm satisfies detailed balance, which is a sufficient condition for its convergence to the desired equilibrium distribution.\cite{frenkel96a}
The requirement of detailed balance can be formulated such that the flow of configurations from an arbitrary state $s_i$ to any other state $s_j$ is equal to flow in the opposite direction,
\begin{equation}
    P_{\ce{obs}}(s_i)P_{\ce{trial}}(s_i \rightarrow s_j)P_{\ce{acc}}(s_i \rightarrow s_j) = P_{\ce{obs}}(s_j)P_{\ce{trial}}(s_j \rightarrow s_i)P_{\ce{acc}}(s_j \rightarrow s_i)
    \label{eq:flow}
\end{equation}
where $P_{\ce{obs}}(s_i)$ is the probability of observing state $s_i$ at equilibrium, $P_{\ce{trial}}(s_i \rightarrow s_j)$ is the probability of proposing a trial move from $s_i$ to $s_j$ and $P_{\ce{acc}}(s_i \rightarrow s_j)$ is the probability of accepting such trial move. 
To satisfy \refeq{eq:flow}, the acceptance probability of our Monte Carlo method must fulfill the following
\begin{equation}
    \frac{P_{\ce{acc}}(s_i \rightarrow s_j)}{P_{\ce{acc}}(s_j \rightarrow s_i)} = \frac{P_{\ce{trial}}(s_j \rightarrow s_i)}{P_{\ce{trial}}(s_i \rightarrow s_j)} \frac{P_{\ce{obs}}(s_j)}{P_{\ce{obs}}(s_i)}.
    \label{eq:detailed_balance}
\end{equation}
In the Reaction ensemble, the probability of observing a reaction state $s$ is
\begin{equation}
    P_{\ce{obs}}(s)=\frac{1}{Q} \prod_x \frac{{(K V^{-1} N_{\ce{A} }^{-1}/\cref)}^{N_x(s)}}{N_x(s)!} \int_V \exp(-\beta U(s,\vec{r}))\mathrm{d}\vec r
\end{equation}
where $Q$ is the partition function in the Reaction ensemble,  $\cref$ is the reference concentration, $U(s,\vec r)$ is the potential energy of the system in the state $s$ in a particular configuration, given by the position vectors of all particles, $\vec r$. 
The product runs over all reacting groups $x$.
In the following, we will only consider $\xi = \pm 1$, meaning that only one bonded pair is being formed ($\xi = +1$) or destroyed ($\xi=-1$) upon transition from state $i$ to $j$.
The number of particles of type $x$ is given by $N_x(s_i)=N_x(s_j)+\xi\nu_x$, where $\nu_x$ is the stoichiometric coefficient of species $x$.
The relative statistical weight of states $s_i$ and $s_j$ follows as
\begin{equation}
  W_{\ce{RxMC}} = \frac{P_{\ce{obs}}(s_j)}{P_{\ce{obs}}(s_i)} = (K V^{-1} N_{\ce{A}}^{-1}/\cref)^\xi \exp{(-\beta \Delta U)} \prod_{x} \frac{N_x(s_i)!}{(N_x(s_i)+\nu_x\xi)!}.
  \label{eq:W_RxMC}
\end{equation}
In the original implementation of the Reaction ensemble Monte Carlo\cite{smith1994} (RxMC), the algorithm proposed the trial moves with equal probability in either direction of the reaction $P_{\ce{trial}}(s_i \rightarrow s_j)=P_{\ce{trial}}(s_j \rightarrow s_i)$. 
Consequently, the acceptance probability was given by Eq. \ref{eq:W_RxMC}. 

In our eb-RxMC algorithm, we introduced an additional bias by sampling of the reaction only within the inclusion radius $\rin$, in order to enable coupling with Langevin Dynamics. 
The proposal probability in our eb-RxMC is asymmetric because the probability of finding two unbonded particles within a given radius is different from that of finding two bonded particles within the same distance.
In the forward direction of the reaction ($\xi=1$), the probability of generating a trial move is given by the probability of finding two unbonded particles within the reaction volume given by $\rin$, 
\begin{equation}
    P_{\ce{trial}}(\xi=+1,r_{\ce{in}}) = P_{\ce{u}} (r_{\ce{in}}) = \frac{\int_0^{r_{\ce{in}}}4\pi r^2 P_{\ce{u}}(r) \ce{d}r}{\int_V 4\pi r^2 P_{\ce{u}}(r) \ce{d}r} 
    \label{eq:prob_unbnd}
\end{equation}
where $P_{\ce{u}}(r)$ is the probability of finding a pair of unbonded particles at a distance $r$. 
Analogously, the probability of performing a trial move in the reverse direction is given by the probability of finding two bonded particles within  $r_{\ce{in}}$,
\begin{equation}
    P_{\ce{trial}}(\xi=-1,r_{\ce{in}}) = P_{\ce{b}}(r_{\ce{in}}) = \frac{\int_0^{r_{\ce{in}}}4\pi r^2 P_{\ce{b}}(r) \ce{d}r}{\int_V 4\pi r^2 P_{\ce{b}}(r) \ce{d}r} 
    \label{eq:prob_bnd}
\end{equation}
where $P_{\ce{b}}(r)$ is the probability of finding a pair of bonded particles at a distance $r$. 
Therefore, the relative statistical weight of the bias introduced by restricting the proposals only within $\rin$ is
\begin{equation}
    W_{\ce{bias}} = \frac{P_{\ce{trial}}(s_i \longrightarrow s_j)}{P_{\ce{trial}}(s_j \longrightarrow s_i)} = \left(\frac{P_{\ce{b}}}{P_{\ce{u}}}\right)^\xi.
    \label{eq:W_biass}
\end{equation}
By substituting Eqs. \ref{eq:W_RxMC} and \ref{eq:W_biass} in Eq. \ref{eq:detailed_balance}, one obtains
\begin{equation}
    \frac{P_{\ce{acc}}(s_i \longrightarrow s_j)}{P_{\ce{acc}}(s_j \longrightarrow s_i)} = \left(\frac{P_{\ce{b}}(r_{\ce{in}})}{P_{\ce{u}}(r_{\ce{in}})} K V^{-1} N_{\ce{A}}^{-1} /\cref \right)^\xi \exp{(-\beta \Delta U)} \prod_{x} \frac{N_x(s_i)!}{(N_x(s_i)+\nu_x\xi)!} = W_{\ce{bias}} W_{\ce{RxMC}} 
\end{equation}
which is our choice as acceptance probability for biased eb-RxMC simulations.

Let us now demonstrate how to calculate $P_{\ce{u}}$ and $P_{\ce{b}}$ for the simplest case of an ideal gas. 
In an ideal gas, the probability of finding a free particle is equal everywhere, hence, \ie $P_{\ce{u}}(r)=\text{const}$, therefore, Eq. \ref{eq:prob_unbnd} simplifies to
\begin{equation}
    P_{\ce{trial}}(\xi=1,r_{\ce{in}})   \overset{\mathrm{ideal}}{=} V_{\ce{R}}(r_{\ce{in}})/V
    \label{eq:P_unbonded_id}
\end{equation}
where $V_{\ce{R}} $ is the reaction volume and $V$ is the volume of the simulation box. 
If the inclusion radius is smaller than half of the simulation box length, the reaction volume is $V_{\ce{R}} = 4/3\pi r_{\ce{in}}^3$. 
However, the inclusion radius can also take higher values, up to the full box length.
In such a case, the reaction volume can be estimated by Monte Carlo integration of the reaction volume inside of the simulation box.
This calculation is an important aspect of validation of our algorithm.
Nevertheless, in practical applications on would typically choose inclusion radius smaller than half of the box length.

\begin{figure}[H]
\centering
\includegraphics [ width =0.99\textwidth]{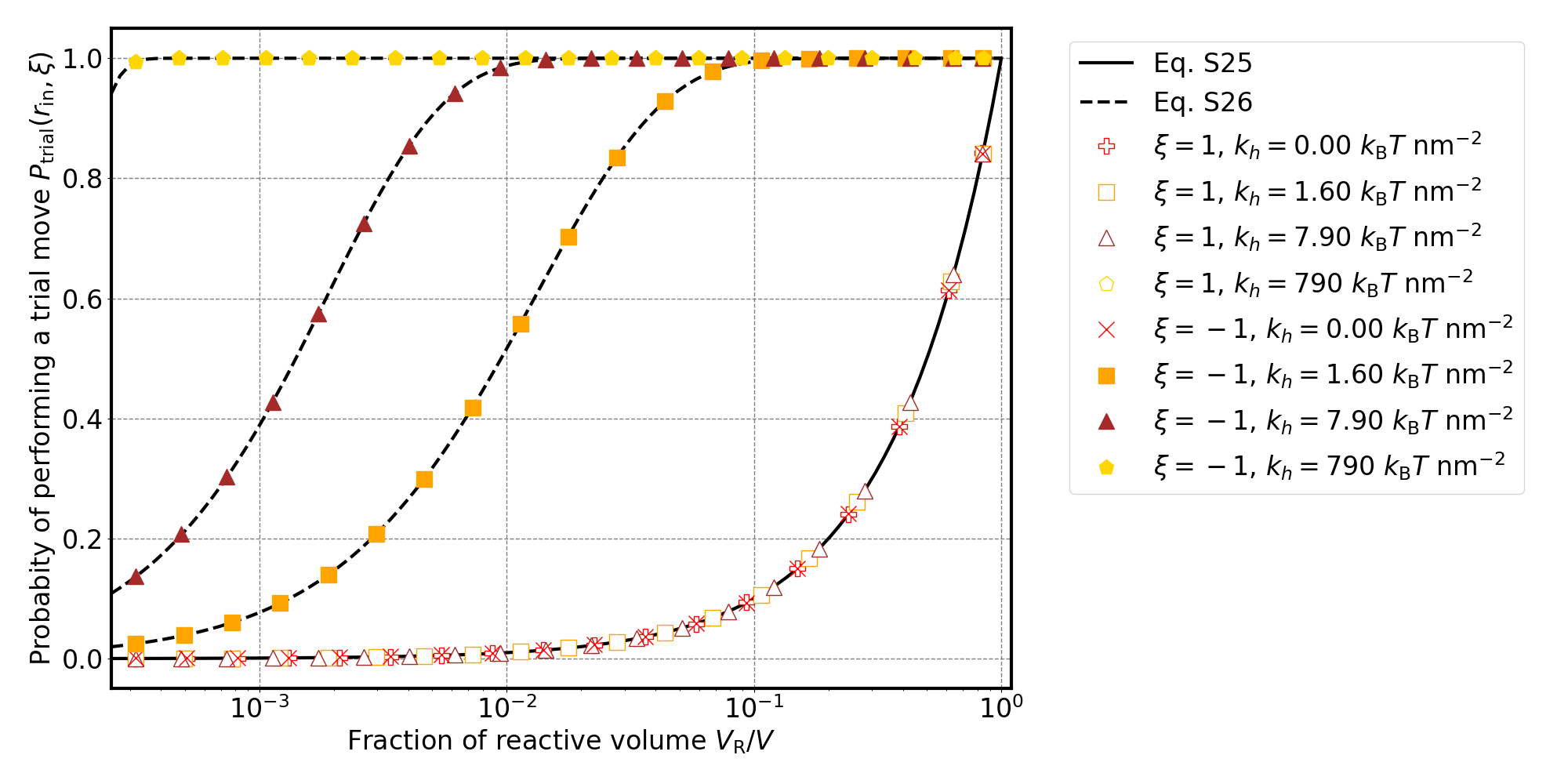}
\caption{Probability of performing a trial move $ P_{\ce{trial}}(\xi,r_{\ce{in}})$ in the forward ($\xi=1$) and reverse ($\xi=-1$) directions of the reaction a function of the fraction of reactive volume $V_{\ce{R}}/V$. 
The markers follow the $ P_{\ce{trial}}(\xi,r_{\ce{in}})$ yielded by our eb-RxMC algorithm at different values of the harmonic constant.
For an ideal gas, $ P_{\ce{trial}}(\xi,r_{\ce{in}})$ can be exactly calculated using  Eqs. \ref{eq:P_bonded_id} (dashed lines) and \ref{eq:P_unbonded_id} (continuous lines). 
}
\label{Fig:probs}
\end{figure}

In the absence of any other interactions, the probability of finding two bonded particles at a particular distance is proportional to the Boltzmann factor, $\exp(-\beta U_{\ce{bond}}(r))$, where $U_{\ce{bond}}(r)$ is the potential energy of the bond and $\beta = 1/(k_{\ce{b}}T)$ is the inverse thermal energy. 
Therefore, one can rewrite Eq. \ref{eq:prob_bnd} as
\begin{equation}
    P_{\ce{trial}}(\xi=-1,r_{\ce{in}})  \overset{\mathrm{ideal}}{=} \frac{\int_0^{r_{\ce{in}}}4\pi r^2 \exp(-\beta U_{\ce{bond}}(r)) \ce{d}r}{\int_V 4\pi r^2 \exp(-\beta U_{\ce{bond}}(r)) \ce{d}r}.
    \label{eq:P_bonded_id}
\end{equation}
One possible choice for the potential energy of the bond is the harmonic potential $U_{\ce{bond}} = \frac{1}{2}k_{\ce{h}}(r-l_0)^2$, where  $k_{\ce{h}}$ is the harmonic constant and $l_0$ is the equilibrium bond length. 
For this choice, the integrals in Eq. \ref{eq:P_bonded_id} have an analytical solution given by
\begin{equation}
     \int_0^{r_{\ce{in}}}4\pi r^2 \exp \left(-\frac{1}{2} \beta k_{\ce{h}}(r-l_0)^2\right) \ce{d}r = 
     A[B(\erf(F_1)+\erf(F_2))-C_1 \exp(E_1)+C_2 \exp(E_2)]
\end{equation}
where $\erf$ is the error function and the coefficients are equal to
\begin{equation}
\begin{aligned}
    &A   = 2\pi (\beta k_{\ce{h}})^{-3/2} \\
    &B   = (\beta k_{\ce{h}} l_0^2+1)\sqrt{2\pi} \\
    &C_1 = 2 (r_{\ce{in}}+l_0) \sqrt{\beta k_{\ce{h}}}  \\
    &C_2 = 2 l_0 \sqrt{\beta k_{\ce{h}}} \\
    &E_1 = -0.5 \beta k_{\ce{h}} (l_0-r_{\ce{in}})^2 \\
    &E_2 = -0.5 \beta k_{\ce{h}} l_0^2 \\
    &F_1 = 2^{-1/2} (r_{\ce{in}}-l_0) \sqrt{\beta k_{\ce{h}}} \\
    &F_2 = 2^{-1/2} l_0 \sqrt{\beta k_{\ce{h}}} \\
\end{aligned}
\end{equation}
We used Eqs.   \ref{eq:P_unbonded_id} and \ref{eq:P_bonded_id} as a self-consistency test to check that our algorithm produces trial moves with the correct probability, irrespective of the choice of $k_{\mathrm{h}}$. 
In Fig. \ref{Fig:probs}, we compare the probabilities of performing a trial move in the forward ($\xi=1$) and reverse ($\xi=-1$) directions yielded by our eb-RxMC simulations (markers) with the expected values given by Eqs. \ref{eq:P_bonded_id} (dashed lines) and \ref{eq:P_unbonded_id} (continuous line) as a function of the fraction of reactive volume $V_{\ce{R}}/V$.
We considered the case of dimerization reactions between chemically identical groups (Eq. \ref{eq:AAbonding}).
We compare simulations with different values of the harmonic constant $k_{\ce{h}}$, which have significantly different trial probabilities in the reverse direction of the reaction. 
In all cases, the agreement between the simulation and the theory is excellent, confirming that our algorithm generates trial configurations with the correct probability. 

\subsection{Sampling the reaction and configuration space of A+B dimerization reactions with $f_\textrm{A}=0.5$}
\begin{figure}[H]
\centering
\includegraphics [ width =0.49\textwidth]{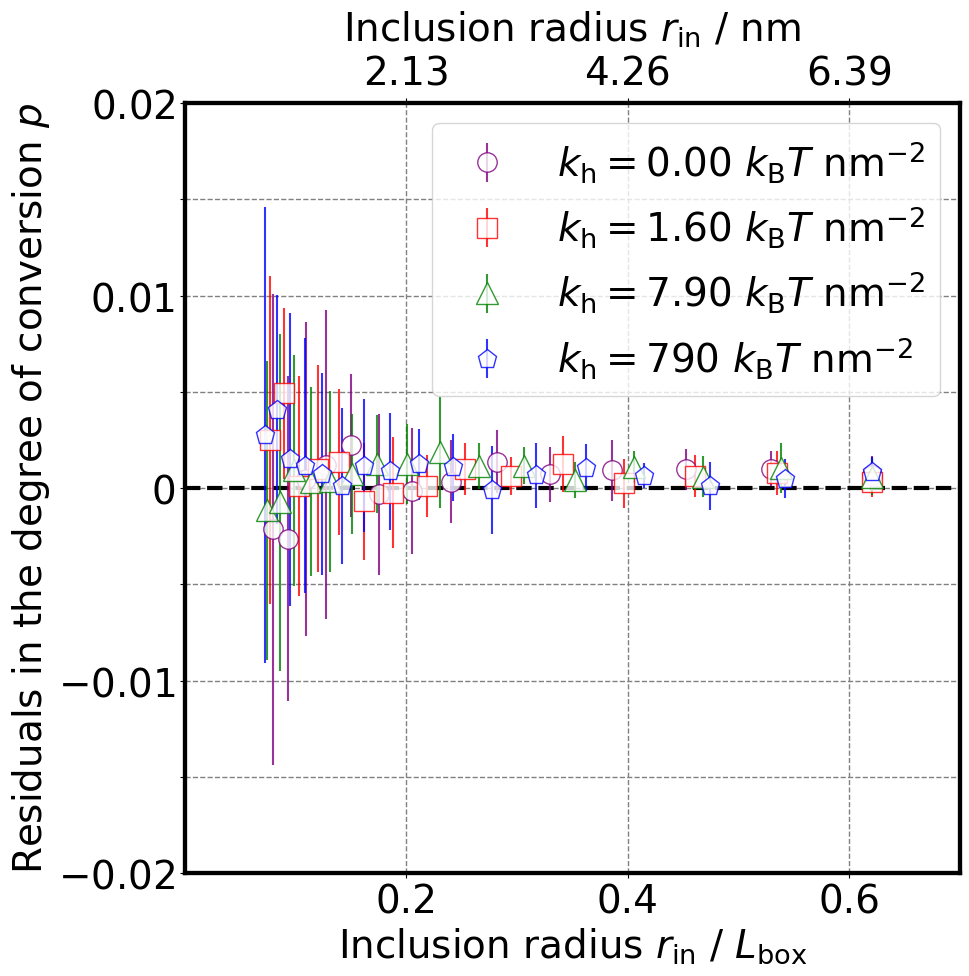}
\includegraphics [ width =0.49\textwidth]{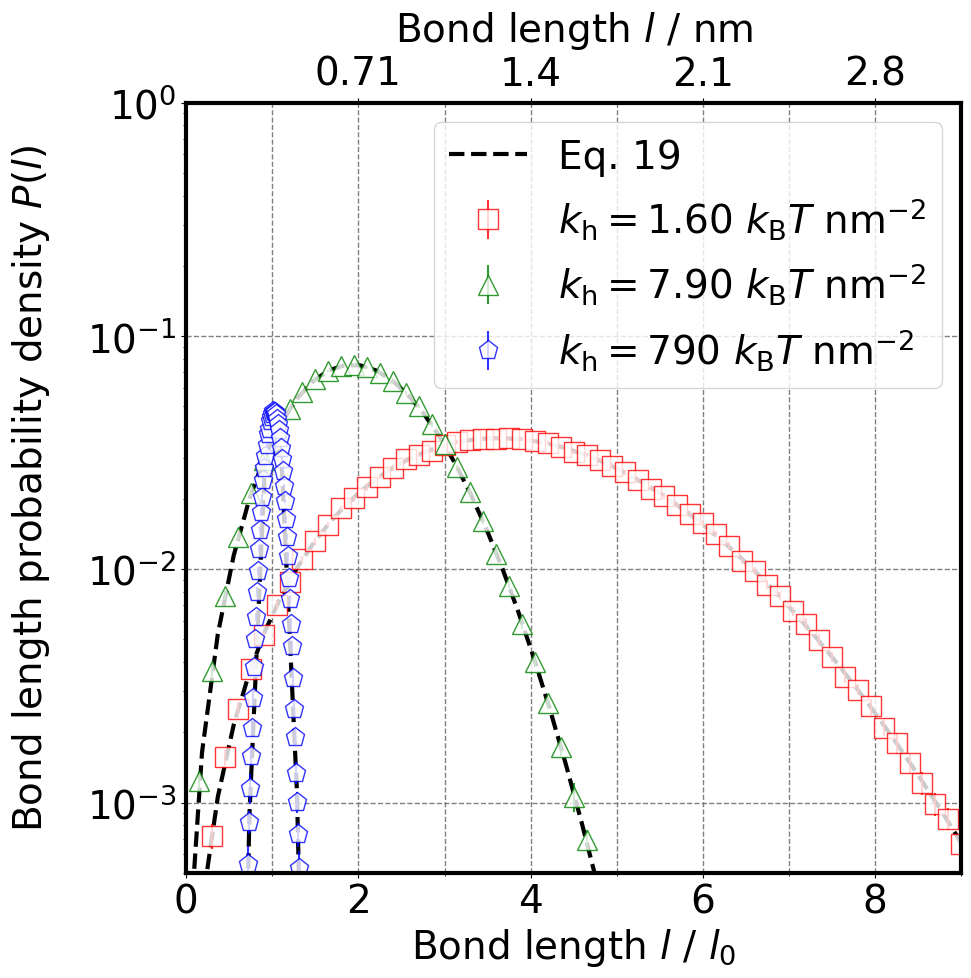}
\caption{Left panel: Residuals in the degree of conversion $p$ as a function of the inclusion radius $\rin$ at various values of the harmonic constant for the reversible bond $\kh$.
The residuals are calculated as the difference between the measured value in our simulations and the expected value $p=0.8$ given by Eq. \ref{eq:p_AB_id} at $\plambda=-1$ and $r_\textrm{ex}=0$, corresponding to $f_\mathrm{A} = 0.5$.
Right panel: Probability density $P(l)$ of observing a reversible bond with a length $l$ in a bin at a distance $r$ with size $\mathrm{d}r$ at different values of $\kh$. 
The numerical results from our simulations (markers) follow the exact result given by Eq. 19 (dashed lines).}
\label{Fig:r_inc_AB}
\end{figure}

\subsection{Finite size effects in eb-RxMC simulation of linear polycondensation}

\begin{figure}[H]
\centering
\includegraphics [ width =0.8\textwidth]{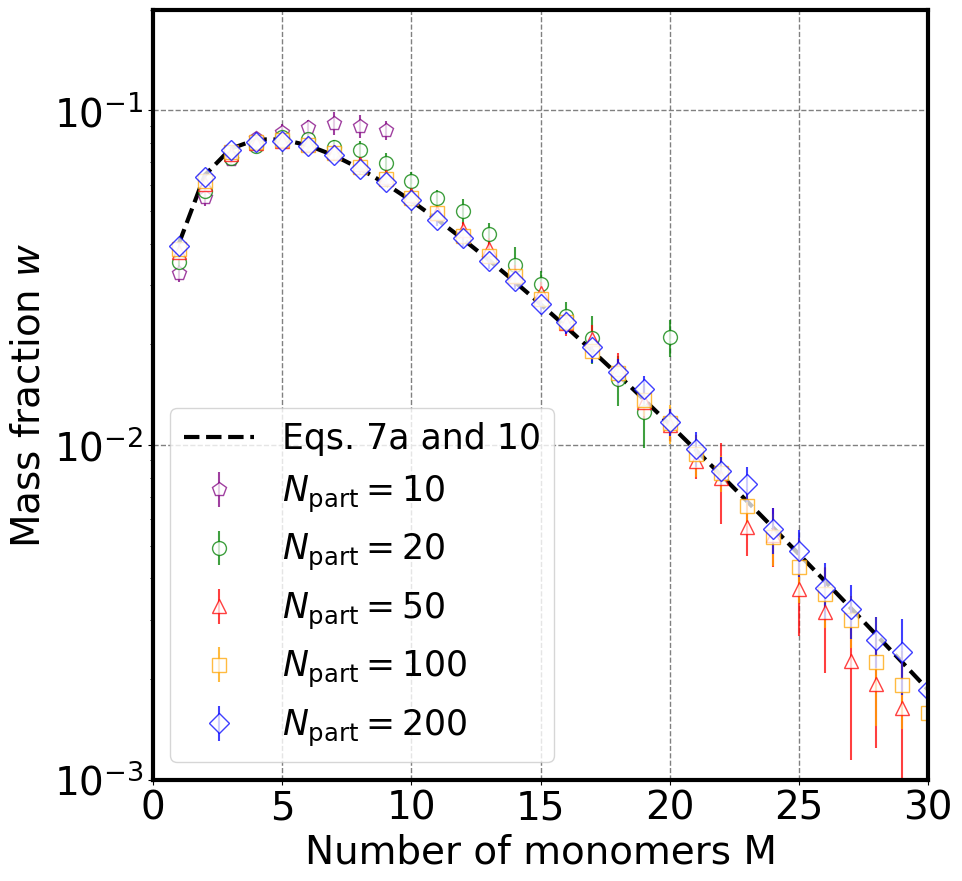}
\caption{Mass fraction $w$ as a function of the number of monomers in the polymer chain $M$ for $\plambda = -1$  at various  values of total number of particles $N_\mathrm{part}$.
 Markers follow the data measured in our eb-RxMC simulation and the dashed line follows the exact value given by Eqs. 7a and 10.
}
\label{fig:finite-size}
\end{figure}

In Fig. \ref{fig:finite-size}, we demonstrate that the mass fraction distribution measured by our eb-RxMC simulation can be affected by artifacts due to finite size effects caused by a relatively small  number of particles $N_{\ce{part}}$ in the simulation box.\cite{hebbeker2023}
To illustrate the effect, let us consider the case of a polycodensation of the type given by Eq.~9b with $\plambda = -1$.
In theory, concentrations of all species are treated as continuous variables.
In contrast, the number of particles of each species in the simulation box can only attain an integer value.
These finite size effects are especially significant if the degree of conversion is very high, so that the most probable chain length is on the same order of magnitude as $N_{\ce{part}}$.
As can be observed for the simulations with lower number of particles $N_{\ce{part}} = 10$ and $N_{\ce{part}} = 20$, the finite size effect causes underestimation of the mass fraction of short chains and, simultaneously, overestimation of the mass fraction of long chains.
Interestingly, the finite size effect does not affect the average degree of conversion, which matches the expected result given by Eq. 7a, at least for the ideal systems studied here.
Nonetheless, for a sufficiently high number of particles, our eb-RxMC simulation correctly sample the distribution of polymer chains matching the exact analytical solution given by Eqs. 7a and 10.

\end{document}

%% file: definitions.tex
\usepackage[version=4]{mhchem}
\usepackage{amsmath}
\usepackage{amssymb}
\usepackage{graphicx}
\usepackage{subfigure}
\usepackage{siunitx}
\usepackage[T1]{fontenc}
\usepackage{xspace}

\usepackage[textsize = small]{todonotes}

\usepackage{xcolor}
\graphicspath{{./figures/}}

\newcommand*{\eg}{\emph{e.g.}\xspace}

\newcommand*{\ie}{\emph{i.e.}\xspace}

\newcommand*{\refeq}[1]{Eq.~\ref{#1}}

\newcommand*{\kb}{k_\mathrm{B}}

\newcommand*{\kT}{k_\mathrm{B}T}

\newcommand*{\pK}{\mathrm{p}K}
\newcommand*{\plambda}{\mathrm{p}\lambda}
\newcommand*{\rin}{r_\mathrm{in}}
\newcommand*{\rex}{r_\mathrm{ex}}
\newcommand*{\pC}{\mathrm{p}C}

\newcommand*{\cref}{c^{\ominus}}
\newcommand*{\ctot}{c_{\ce{tot}}}
\newcommand*{\cmax}{c^{\ce{max}}}
\newcommand*{\Pb}{P_{\ce{b}}}
\newcommand*{\Pu}{P_{\ce{u}}}
\newcommand*{\kh}{k_{\ce{h}}}
\newcommand*{\Npart}{N_\mathrm{part}}

\DeclareMathOperator\erf{erf}

%% file: head.tex
\title{The explicit bonding Reaction ensemble Monte Carlo method}

\newcommand*{\affCUNI}{Department of Physical and Macromolecular Chemistry, Faculty of Science, Charles University, Hlavova 8, 128 40 Prague 2, Czech Republic}
\newcommand*{\affIQTCUB}{Department of Material Science and Physical Chemistry, Research Institute of Theoretical and Computational Chemistry (IQTCUB), University of Barcelona, Martí i Franquès 1, 08028 Barcelona, Spain}
\newcommand*{\affNTNU}{Department of Physics, NTNU - Norwegian University of Science and Technology, NO-7491 Trondheim, Norway}

\author{Pablo M. Blanco}
\email{pablb@ntnu.no}
\affiliation{\affCUNI}
\alsoaffiliation{\affIQTCUB}
\alsoaffiliation{\affNTNU}

\author{Peter Košovan}
\affiliation{\affCUNI}
\email{peter.kosovan@natur.cuni.cz}

\keywords{Reaction ensemble, polycondensation}
